\documentclass[aps,11pt,preprintnumbers,nofootinbib,floatfix]{revtex4}

\usepackage{graphicx}
\usepackage{epsfig}
\usepackage{dcolumn}
\usepackage{subfigure}
\usepackage{multirow}
\usepackage{amsmath}
\usepackage{amssymb}

\begin{document}

\title{Entanglement entropy and Page curve of black holes with island in massive gravity}


\author{Cao H. Nam}
\email{nam.caohoang@phenikaa-uni.edu.vn}  
\affiliation{Phenikaa Institute for Advanced Study and Faculty of Fundamental Sciences, Phenikaa University, Yen Nghia, Ha Dong, Hanoi 12116, Vietnam}
\date{\today}

\begin{abstract}%
By applying the island rule proposed recently, we compute the entanglement entropy of Hawking radiation and study the Page curve for the eternal black holes in massive gravity. We investigate for both the neutral and charged black holes which the corresponding results of Schwarzschild and Reissner-Nordstr\"{o}m black holes are restored in the limit of massless graviton. We show for the neutral and non-extremal charged black holes that the island is not formed at the early times of the evaporation and hence the entanglement entropy increases linearly in time. However, for the extremal charged black hole, the calculation of the entanglement entropy at the early times without the island is ill-defined because the metric is divergent at the curvature singularity. This implies that new physics in the UV region must be taken into account to make the metric behaving smoothly at the very short distances. At the late times, with the emergence of one island near the event horizon, the entanglement entropy is saturated by the Bekenstein-Hawking entropy of black holes. In addition, we analyze the impact of massive gravity parameters on the size of island, the entanglement entropy, the Page time, and the scrambling time in detail.
\end{abstract}

\maketitle

\section{Introduction}

The behavior of back holes as the thermodynamic objects with the temperature and entropy (which are determined in terms of the surface gravity and the horizon area, respectively) provides a deep connection between the research areas of general relativity, thermodynamics, and quantum mechanics \cite{Bekenstein1972,Bekenstein1973,Bekenstein1974,Bardeen1973,Hawking1975}. In addition, the black hole thermodynamics may offer indispensable insights into quantum gravity. By taking into account the quantum effects of the matter fields near the event horizon, S. Hawking showed that black holes can emit the radiation with the nearly thermal spectrum \cite{Hawking1975}. Suppose that black holes are formed by the gravitational collapse of the matter with the initial state in a pure (quantum) state, after black holes evaporate completely the final state of the system would be in a mixed (thermal) state. In this way, the quantum information of the initial state is not conserved in the evaporation process of black holes, leading to the so-called information loss paradox \cite{Hawking1976}. On the other hand, the time evolution in this way is contradictory to one of the fundamental principles of quantum mechanics, namely the unitarity principle which requires that the final state must be the pure state if the system starts from the same kind of the state.

\begin{figure}[t]
 \centering
\begin{tabular}{cc}
\includegraphics[width=0.5 \textwidth]{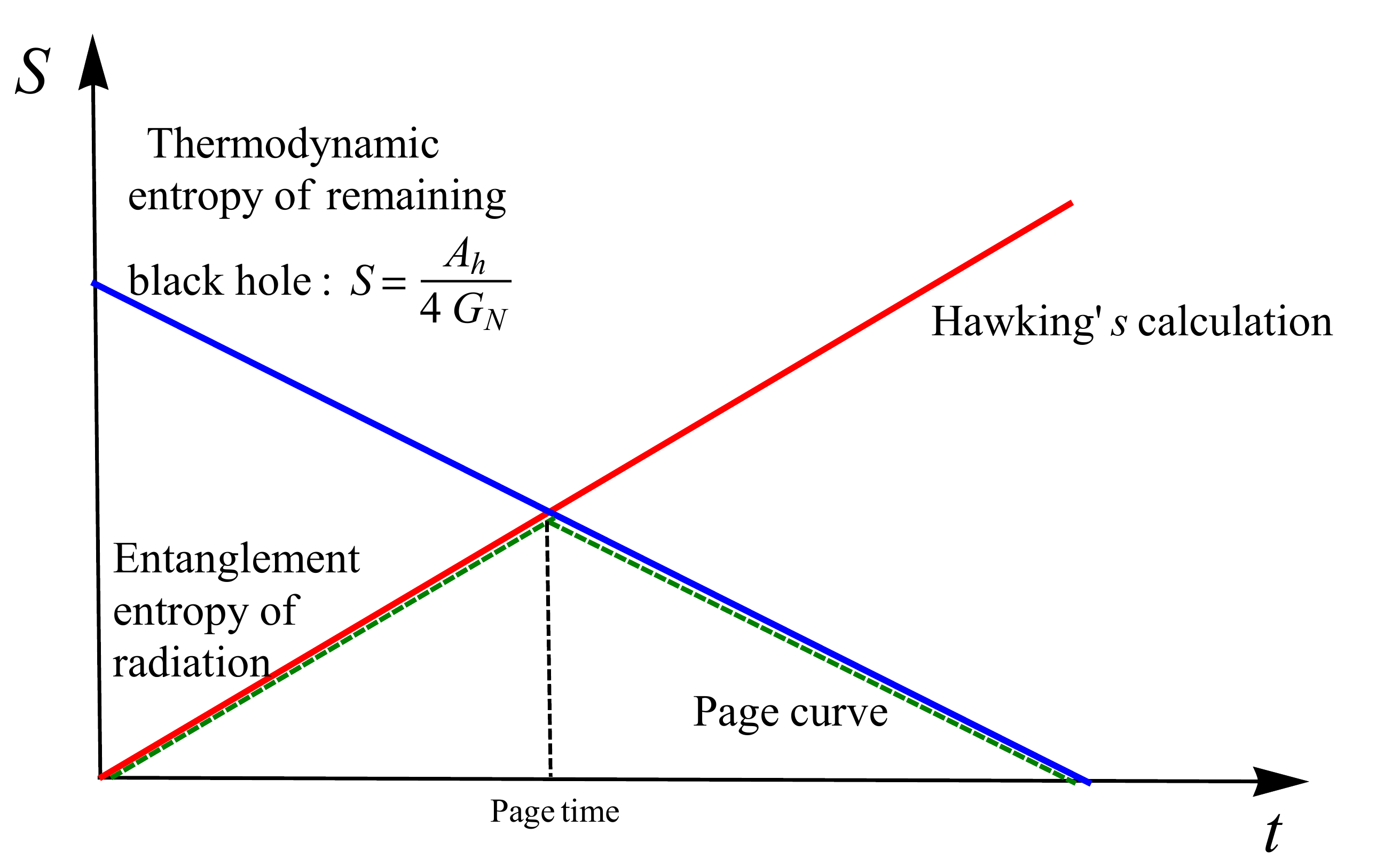}
\end{tabular}
 \caption{The Page curve, described by the green dashed line, for the black hole evaporation consistent with the unitarity evolution.}\label{BH-Page-curve}
\end{figure}

If a black hole is formed from a pure state, then its entropy is zero. Later, the black hole would evaporate due to emitting the Hawking radiation. During the early time of the black hole evaporation, the entanglement entropy of Hawking radiation should increase in time because more and more Hawking quanta are emitted and entangled with the remaining black hole. Whereas, the thermodynamic or Bekenstein-Hawking entropy of the black hole would decrease due to its horizon area shrinking. The behavior of the entanglement entropy of Hawking radiation changes at the time when the Bekenstein bound is violated. The Bekenstein bound implies that the fine-grained entropy of the black hole should not be larger than the Bekenstein-Hawking entropy of the black hole.  As the degrees of freedom of the remaining black hole together with the outgoing radiation as a whole is pure state, the fine-grained entropy of the remaining black hole should be equal to the entanglement entropy of the Hawking radiation. Hence, at the moment at which the entanglement entropy of Hawking radiation is equal to the Bekenstein-Hawking entropy of the remaining black hole, called the Page time, the entanglement entropy of Hawking radiation must decrease and drop down to zero as the black hole evaporates completely. This means that the quantum information of the initial black hole is encoded in the Hawking radiation and the black hole evaporation is consistent with the unitarity evolution. In this way, the unitarity evolution of the black hole evaporation corresponds to that the entanglement entropy of Hawking radiation follows the Page curve \cite{Page1993a,Page1993b}: the entanglement entropy of Hawking radiation increases monotonically at the early period of the black hole evaporation, then reaches a maximum value at the Page time, and finally has to go to zero at the end of the evaporation process, as indicated in Fig. \ref{BH-Page-curve}. 
It is expected that the Page curve is obtained in a theory of quantum gravity. Therefore, reproducing the Page curve for the time evolution of the black hole evaporation is an important step towards not only resolving the information loss paradox of black holes but also understanding completely quantum gravity.

The recent works have shown a significant progress in deduducing the Page curve for the entanglement entropy of Hawking radiation by taking into account the configuration with the islands \cite{Engelhardt2015,Almheiri2019,Penington2020,Mahajan2019}. The islands are some regions $I$ which are completely disconnected from the region $R$ of Hawking radiation which is assumed to be far away from black holes such that the backreaction of Hawking quanta on the spacetime geometry is negligible. The boundaries of the islands extremize the generalized entropy functional and hence they are called the extremal surfaces. A density matrix relating to the states of Hawking radiation is normally calculated by taking the partial trace over the states in the complementary part of the radiation region. However, by using the quantum extremal surface technique, it was found that the islands appear in the complementary region of $R$ at the late times of the black hole evaporation process and hence the states in the islands should be eliminated from the states which are traced out. According to the quantum extremal surface prescription, the entanglement entropy of Hawking radiation is obtained as the minimum value of the generalized entropy functional \cite{Engelhardt2015,Ryu2006,Hubeny2007,Akers2020,Faulkner2013,Wall2014}
\begin{eqnarray}
S(R)=\text{min}\left\{\text{ext}\left[\frac{\mathcal{A}(\partial I)}{4G_N}+S_{\text{mat}}(R\cup I)\right]\right\},\label{gen-ent}
\end{eqnarray}
where $G_N$ is the Newton gravitational constant, $\partial I$ refers to the island boundaries, $\mathcal{A}(\partial I)$ is the total area or volume of the island boundaries for $D=4$ or $D>4$, respectively, and $S_{\text{mat}}$ is the von Neumann entropy of the quantum fields on the union of the radiation and island regions. 

\begin{figure}[t]
 \centering
\begin{tabular}{cc}
\includegraphics[width=0.5 \textwidth]{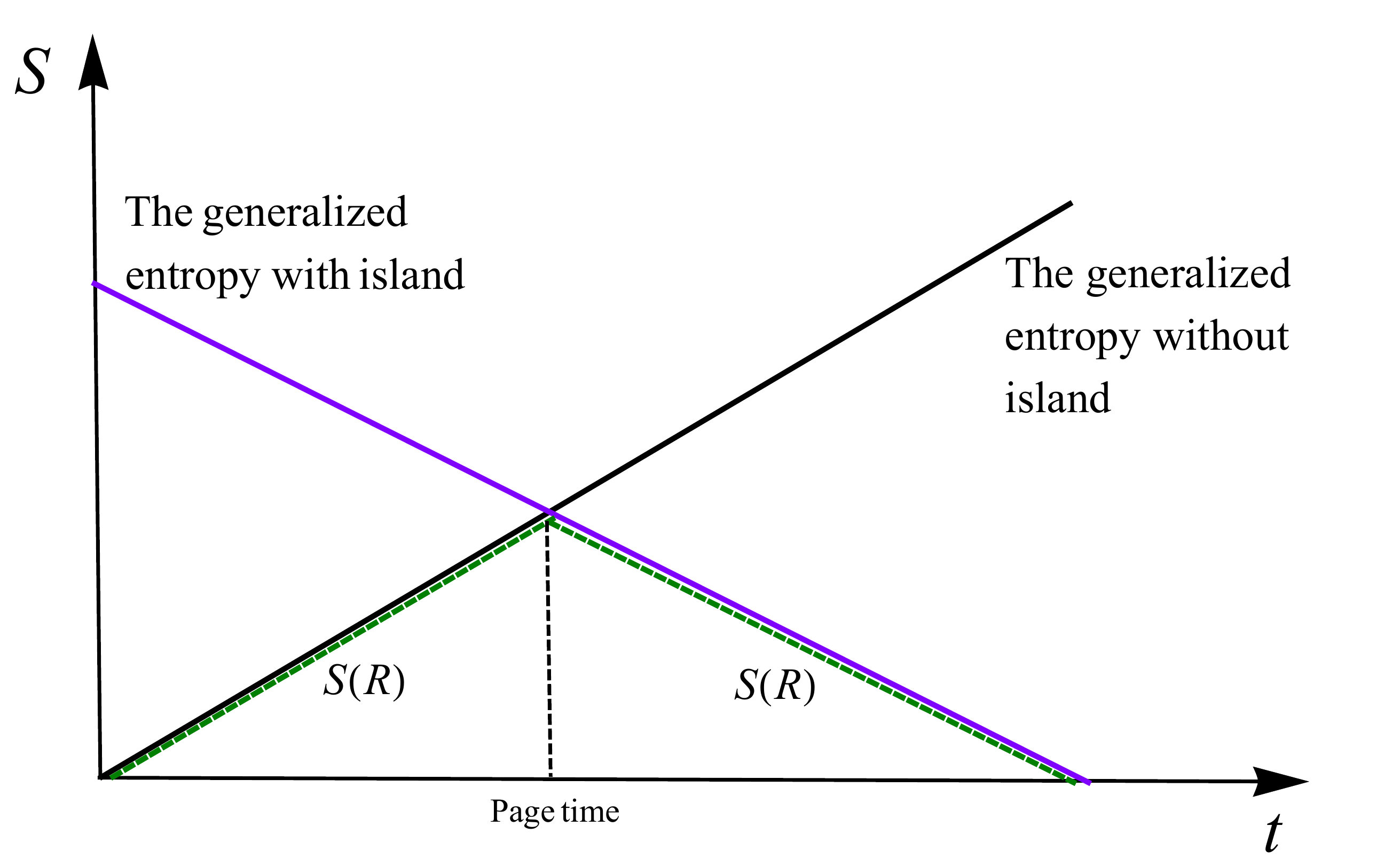}
\end{tabular}
 \caption{The Page curve, described by the green dashed line, for the black hole evaporation with the island method.}\label{PC-island}
\end{figure}

The island formula (\ref{gen-ent}) means that the entanglement entropy of Hawking radiation is obtained as the minimum value of the generalized entropy functional over all possible extremal surfaces corresponding to all possible locations of the island. The entanglement entropy of Hawking radiation calculated in this method contains two contributions which come from the area term of the island and the von Neumann entropy of the quantum fields on the union of the radiation and island regions. At the early time of the black hole evaporation, there is no island forming and since the black hole entanglement wedge contains all the black hole interior. The entanglement entropy of Hawking radiation thus is the von Neumann entropy of quantum fields in the black hole exterior and it increases as more and more Hawking quanta are emitted. However, at the late time, there is the emergence of the island whose boundary is very close the black hole horizon and which extends almost through the whole black hole interior. Consequently, the partners of Hawking quanta which fall inside the black hole are almost contained in the island. This means that the von Neumann entropy of the quantum fields on the union of the radiation and island regions is small. On the other hand, the dominant contribution for the entanglement entropy of Hawking radiation is from the area term of the island. When the black hole shrinks due to the evaporation, this term would decrease. Therefore, the entanglement entropy of Hawking radiation would go to zero when the black hole evaporates completely. This means that the behavior of the entanglement entropy of Hawking radiation with the island method follows the Page curve, as depicted in Fig. \ref{PC-island}.

Interestingly, it was shown that the configuration with the island can be emerged from the gravitational Euclidean path integral using the replica trick \cite{Hartman2020,Stanford2019}. In the gravitational replica method, the different replica sheets with the boundary conditions kept fixed are connected together by the so-called replica wormholes which are new saddle points in the gravitational Euclidean path integral. In the semi-classical limit, they are the dominant contributions to the effective action for the geometry which is a sum of the gravitational action and the partition function of the quantum fields. In this way, the presence of the replica wormholes yields the same island formula for the entanglement entropy of Hawking radiation as using the quantum extremal surface technique.

The island rule was initially considered for the two-dimensional gravitational systems where the explicit computations for the entanglement entropy of Hawking radiation and the Page curve can be easily performed by using the semi-classical method due to the presence of the high symmetries. For the two-dimensional black holes in the context of Jackiw-Teitelboim (JT) gravity, the islands are emerged at the later times of the black hole evaporation and hence their presence makes the entanglement entropy of Hawking radiation remaining finite at the late stage of the evaporation process \cite{Mahajan2019,Hartman2020}. The island consideration in the two-dimensional models was extended to study the asymptotically flat 2d dilaton black holes \cite{Anegawa2020}, the two-sided Janus black holes \cite{Bak2021}, and the evaporating black holes \cite{Goto2021,Hollowood2020,WangLi2021}.

Although the four- and higher-dimensional gravitational systems are complicated due to the lack of the symmetric analyses, the recent works have demonstrated that the island rule is applicable to calculate the entanglement entropy of Hawking radiation and reproduce the Page curve consistent with the unitarity time evolution. For the four- and higher-dimensional eternal Schwarzschild black holes, the authors in \cite{Hashimoto2020} showed the emergence of an island whose dominant contribution leads to the finiteness of the entanglement entropy of Hawking and the Page curve incorporating the unitarity principle. Here, the boundaries of the island are located in the outer vicinity of the event horizon of the Schwarzschild black holes. For considering the backreaction of Hawking radiation, which is natural in the context of black holes in a cosmology supported by the radiation, the existence of islands was pointed out in the cosmological braneworlds \cite{Krishnan2021}. These results have also been confirmed for the Reissner-Nordstr\"{o}m (RN) black holes \cite{LiWang2021,KimNam2021}, the charged/neutral dilaton black holes \cite{Karananas2021,YuGe2021,Ahn2021}, the Kaluza-Klein black holes \cite{LuLin2021}, and the black holes including the higher derivative terms \cite{Alishahiha2021}. Additionally, the islands corresponding to the left/right entanglement of a conformal defect was studied in Randall-Sundrum braneworld model involving weakly gravitating bath \cite{Bhattacharya2021}. In this direction, there are also the investigations about the entanglement of purification and complexities for multi-boundary wormhole models of islands \cite{ABhattacharya2020,Maulik2021}.

As mentioned, the island rule has been extended to calculate the entanglement entropy of Hawking radiation for various black hole geometries which are well-known in the literature for the four- and higher-dimensional cases. But, graviton is massless in the gravity frameworks which are equipped to derive these black hole geometries. In addition, it was argued that the island proposal coupling the gravitating bath induces a mass for the bulk graviton \cite{Hgeng2020}. Whereas, the authors in \cite{HaoGeng2021} argued that the islands might not constitute the consistent entanglement wedges in the gravity theories with massless graviton. These results imply that calculating the entanglement entropy of Hawking radiation and the Page curve for black holes using the island method would be in the context of massive gravity. 

Considering a nonzero mass of graviton is one of the infrared (IR) modifications of gravity, which has the cosmological consequences of which the lately accelerating expansion of the universe can be naturally explained without invoking dark energy.\footnote{Recently, an upper bound on the mass of graviton as $m_g<1.2\times10^{-22}$ eV has been derived from the direct observations of the gravitational waves by LIGO \cite{LIGO2016}.} There are the first attempts to construct a gravity theory accompanied with a nonzero mass of graviton, such as Fierz-Pauli (linear) massive gravity \cite{Fierz1939} or nonlinear massive gravity with the Vainshtein mechanism \cite{Vainshtein1972}. However, these  massive gravity theories suffer from the pathological problems which are well-known as the van Dam-Veltman-Zakharov discontinuity \cite{Dam1970,Zakharov1970} (the predictions in the massless limit do not coincide with those of Einstein gravity) and the Boulware-Deser (BD) ghost \cite{Boulware1972}. These pathological problems thus became the obstacle in establishing a consistent theory of massive gravity. Until recent years, a significant progress has been made by de Rham, Gabadadze and Tolley (dRGT) who proposed a nonlinear massive gravity which is free of the BD ghost and gives the same prediction in the massless limit as Einstein gravity does \cite{deRham2010,deRham2011}. 
It was shown that the ghost-free potential structure of massive gravity does not change under the quantum corrections at one-loop \cite{Heisenberg2013}. In the metric formulation of massive gravity, it is difficult to write down the interactions of different helicity modes of massive spin-2 field due to the square root structure of potential. However, using the vierbein language one can obtain the decoupling limit of massive gravity which consists of the full interactions of the helicity-1 and helicity-0 modes \cite{Ondo2013}. For reviews about the recent progress in massive gravity, see \cite{deRham2014,Hinterbichler2012}.

It was also pointed that massive gravity has the interesting implications for cosmology. The authors in \cite{Mukohyama2011} argued that massive gravity allows open Friedmann-Robertson-Walker (FRW) cosmological solutions  contrary to the no-go result \cite{Dubovsky2011} which does not extend to the open FRW universes. Perturbations around FRW background solutions were studied with the extra metric which is promoted to be dynamical \cite{Crisostomi2012}. New massive gravity theories proposed in \cite{Kimura2020} contain stable cosmological solutions without the requirement of infinite strong coupling. The approaches towards healthy cosmological solutions were proposed in \cite{Felice2013}. In addition, various black hole solutions and their thermodynamic properties have been extensively investigated in massive gravity \cite{Gao2013,Cai2015,Ghosh2016,Prasia2016,Panahiyan2016c,SFernando2016,Zou2017,Tannukij2017,Guo2017,Yue2017,Boonserm2018,
Nam2018,Amirabi2020,Singh2020,SQHu2020,Nam2020,Khan2020,Sadeghi2021}. The novel thermodynamic phenomena of black holes in the extended phase space with the cosmological constant identified as the thermodynamic pressure have been pointed out in the context of massive gravity like van der Waals phase transition of black holes \cite{Mann2017}, the efficiency of heat engine provided by black holes \cite{Panahiyan2018,Nam2021}, and Joule-Thomson expansion of black holes \cite{CHNam2020}. The entanglement entropy in the van der Waals phase transition in the context of massive gravity was studied in \cite{LiuNiu2021}. For another massive object, the the structure of neutron star was investigated in the context of massive gravity \cite{Bordbar2017}. Furthermore, the effects of massive gravity on s-/p-wave holographic superconductors have been explored in \cite{HZeng2020,HNam2020,QXiang2020}. 

Because of many interesting aspects of massive gravity, it is worth to study the entanglement entropy of Hawking radiation and the Page curve by applying the island rule for the black hole geometries in the context of massive gravity and analyze the effects due to the nonzero mass of graviton on them. We consider the four-dimensional black hole solutions in massive gravity which were found in \cite{Cai2015}. The presence of graviton mass in massive gravity leads to two new terms in the black hole solutions. The first term corresponds to a linear term in the radial coordinate $r$ which would play the role of the quintessence matter with the proper coupling parameter of massive gravity. Whereas, the second term corresponds to the global monopole solution which was obtained in the presence of topological defect as a result of the spontaneous symmetry breaking in the early universe \cite{Barriola1989}. Interestingly, the presence of the linear term leads to the asymptotically non-flat behavior of black hole geometries. The entanglement entropy of Hawking radiation and the Page curve are explored by the island method for various black hole geometries corresponding to the asymptotically flat (AdS/dS) behavior. Hence, it is worth checking whether the island method is still able to be applied to the black hole geometries with the asymptotically non-flat behavior, which is important to show the wide range of applicability of the island method. In addition, the linear term and global monopole term modify the dynamic properties and the evaporation of black holes. Thus, these terms would lead to the changes in the entanglement entropy of Hawking radiation and the Page curve. We find the expressions for the entanglement entropy of Hawking radiation, the Page time, and the scrambling time in the presence of linear term and global monopole term and analyze their impacts on these quantities. We point to that in the limit of massless graviton the expressions for the entanglement entropy of Hawking radiation, the Page time, and the scrambling time would reduce to the corresponding expressions in the context of Einstein gravity. This shows a continuity of dRGT massive gravity with Einstein gravity in the limit which the graviton mass approaches zero and thus supports theory of dRGT massive gravity as the consistent theory of modified gravity. 

This paper is organized as follows. In Sec. \ref{BHinMG}, we introduce the four-dimensional black hole solutions in massive gravity which were found in \cite{Cai2015}. Then, we determine various related coordinates and rewrite the metric in the Kruskal coordinate, which are all basis for the evaluations in the next sections. In Sec. \ref{EEforNBH}, we calculate the entanglement entropy of Hawking radiation for the neutral black hole without and with the island which corresponds to the early and late times of the black hole evaporation, respectively. In addition, we derive the Page and scrambling times and study the effects of massive gravity on these quantities, the location of the island boundaries, and the entanglement entropy of Hawking radiation. In Sec. \ref{EEforCBH}, we repeat the calculations of Sec. \ref{EEforNBH} for the charged black hole in massive gravity for the non-extremal and extremal cases. With respect to the non-extremal case, the effects of massive gravity on the Page time and the scrambling time are qualitatively the same as the neutral black hole when the distance between the event and Cauchy horizons is not so close, but in the contrary they drastically change. The results found for the extremal case are almost different from the one of the neutral and non-extremal charged black holes due to their causal structure. Finally, we conclude in Sec. \ref{conclu}.

\section{\label{BHinMG}Black holes in massive gravity}
In this section, we review briefly the black hole solutions in the framework of massive gravity in four dimensions, found in \cite{Cai2015}. Then, we introduce the Kruskal coordinate and rewrite the line element in this coordinate for various black hole geometries in massive gravity. 

The action of massive gravity coupled to the Maxwell field in four dimensions is given by
\begin{eqnarray}
S=\frac{1}{16\pi G_N}\int
d^4x\sqrt{-g}\left[R+m^2_g\sum^4_{i=1}c_i\mathcal{U}_i(g,f)-\frac{1}{4}F^{\mu\nu}F_{\mu\nu}\right],
\end{eqnarray}
where $R$ is the scalar curvature of spacetime, $m_g$ is the graviton mass, $c_i$ are the coupling parameters, $f$ is the reference (fiducial) metric which is not dynamical, $\mathcal{U}_i$ are symmetric polynomials which are written in terms of the eigenvalues of the $4\times4$ matrix
${\mathcal{K}^\mu}_\nu=\sqrt{g^{\mu\lambda}f_{\lambda\nu}}$ as
\begin{eqnarray}
\mathcal{U}_1&=&[\mathcal{K}],\nonumber \\
\mathcal{U}_2&=&[\mathcal{K}]^2-[\mathcal{K}^2],\nonumber \\
\mathcal{U}_3&=&[\mathcal{K}]^3-3[\mathcal{K}][\mathcal{K}^2]+2[\mathcal{K}^3],\nonumber\\
\mathcal{U}_4&=&[\mathcal{K}]^4-6[\mathcal{K}]^2[\mathcal{K}^2]+8[\mathcal{K}][\mathcal{K}^3]+3[\mathcal{K}^2]^2-6[\mathcal{K}^4],\nonumber\label{masgrav-pols}
\end{eqnarray}
with $[\mathcal{K}]={\mathcal{K}^\mu}_\mu$. The corresponding equations of motion read
\begin{eqnarray}
G_{\mu\nu}+m^2_g\chi_{\mu\nu}&=&\frac{1}{2}F_{\mu\lambda}{F_\nu}^\lambda-\frac{1}{8}g_{\mu\nu}F^{\rho\lambda}F_{\rho\lambda},\nonumber\\
\nabla_\mu F^{\mu\nu}&=&0,
\end{eqnarray}
where 
\begin{eqnarray}
\chi_{\mu\nu} &=&-\frac{c_{1}}{2}\left(\mathcal{U}_{1}g_{\mu\nu}-\mathcal{K}_{\mu\nu}\right)-\frac{c_{2}}{2}\left(\mathcal{U}_{2}g_{\mu\nu }-2\mathcal{U}_{1}\mathcal{K}_{\mu\nu}+2\mathcal{K}_{\mu\nu}^{2}\right)-\frac{c_{3}}{2}(\mathcal{U}_{3}g_{\mu\nu}-3\mathcal{U}_{2}\mathcal{K}_{\mu\nu}\nonumber\\
&&+6\mathcal{U}_{1}\mathcal{K}_{\mu\nu}^{2}-6\mathcal{K}_{\mu\nu}^{3})-\frac{c_{4}}{2}(\mathcal{U}_{4}g_{\mu \nu}-4\mathcal{U}_{3}\mathcal{K}_{\mu\nu}+12\mathcal{U}_{2}\mathcal{K}_{\mu\nu}^{2}-24\mathcal{U}_{1}\mathcal{K}_{\mu\nu}^{3}+24\mathcal{K}_{\mu\nu}^{4}).
\end{eqnarray}
With the choice of gauge-fixed ansatz for the reference metric as
\begin{eqnarray}
f_{\mu\nu}=\text{diag}(0,0,c^2_0,c^2_0\sin^2\theta),
\end{eqnarray}
where $c_0$ is a constant set to be one without loss of generality, the spherically symmetric black hole solution is found as \cite{Cai2015}
\begin{eqnarray}
ds^2&=&-f(r)dt^2+\frac{dr^2}{f(r)}+r^2d^2\Omega_{2},\nonumber\\
F_{\mu\nu}&=&\left(\delta^t_\mu\delta^r_\nu-\delta^t_\nu\delta^r_\mu\right)\phi(r),
\end{eqnarray}
where
\begin{eqnarray}
f(r)&=&1-\frac{2M}{r}+\frac{Q^2}{r^2}+m^2_g\left(\frac{c_1r}{2}+c_2\right),\nonumber\\
\phi(r)&=&\frac{Q}{r^2},\label{fr-phir-exp}
\end{eqnarray}
with $M$ and $Q$ to be the ADM mass and the electric charge of the black hole. We see that this black hole solution would reduce to the Schwarzschild black hole in the massless limit $m_g\rightarrow0$. The asymptotic behavior of the black hole solution is dependent on the sign of the coupling parameter $c_1$. For $c_1<0$, the term $m^2_gc_1r/2$ in the metric function $f(r)$ plays the role of the quintessence matter with the quintessential state parameter $\omega_q=-2/3$ \cite{Carroll1998,Copeland2006}. In this situation, besides the solutions of the black hole horizons, the equation $f(r)=0$ leads to the solution of the cosmological horizon. In this work, we only consider the positive sign of $c_1$ which corresponds to the absence of the cosmological horizon.

In the following, we introduce the Kruskal coordinate and rewrite the line element in this coordinate for two cases: (i) Neutral black hole and (ii) Charged black hole.

\subsection{Neutral black hole}
As the electric charge of the black hole is zero, i.e. $Q=0$, we obtain the neutral black hole solution. In this case, the black hole only has one (event) horizon which is denoted by $r_h$ given by
\begin{eqnarray}
r_h=\frac{1}{m^2_gc_1}\left[\sqrt{4Mm^2_gc_1+(1+m^2_gc_2)^2}-(1+m^2_gc_2)\right].
\end{eqnarray}
We can rewrite the metric function $f(r)$ in the event horizon $r_h$ as follows
\begin{eqnarray}
f(r)=\frac{m^2_gc_1}{2}\frac{(r-r_h)(r+r_h+\delta)}{r},
\end{eqnarray}
where 
\begin{eqnarray}
\delta\equiv\frac{2(1+m^2_gc_2)}{m^2_gc_1}.
\end{eqnarray}
By defining the tortoise coordinate as
\begin{eqnarray}
r_*=\int\frac{dr}{f(r)}=\frac{1}{2\kappa}\left[\log\frac{|r-r_h|}{r_h}+\left(1+\frac{\delta}{r_h}\right)\log\left(\frac{r+r_h+\delta}{r_h+\delta}\right)\right],
\end{eqnarray}
where $\kappa$ refers to the surface gravity at the event horizon given by
\begin{eqnarray}
\kappa=\frac{f'(r_h)}{2}=\frac{m^2_gc_1(2r_h+\delta)}{4r_h},
\end{eqnarray}
we introduce the Eddington-Finkelstein coordinate constructed relying on the paths of the radially incoming and outgoing photons as follows
\begin{eqnarray}
u=t-r_*,\ \ \ \ v=t+r_*.
\end{eqnarray}
Then, the Kruskal coordinate is defined as
\begin{eqnarray}
U=-e^{-\kappa u}, \ \ \ \ V=e^{\kappa v}.
\end{eqnarray}
The line element is rewritten in terms of the Kruskal coordinate as
\begin{eqnarray}
ds^2=-W^2(r)dUdV+r^2d\Omega^2_2,\label{Krcood-NBH}
\end{eqnarray}
where the conformal factor $W^2(r)$ is given by
\begin{eqnarray}
W^2(r)=\frac{8}{m^2_gc_1}\frac{r^3_h(r_h+\delta)}{r(2r_h+\delta)^2}\left(\frac{r+r_h+\delta}{r_h+\delta}\right)^{-\frac{\delta}{r_h}}.
\end{eqnarray}

\subsection{Charged black hole}
\begin{figure}[t]
 \centering
\begin{tabular}{cc}
\includegraphics[width=0.5 \textwidth]{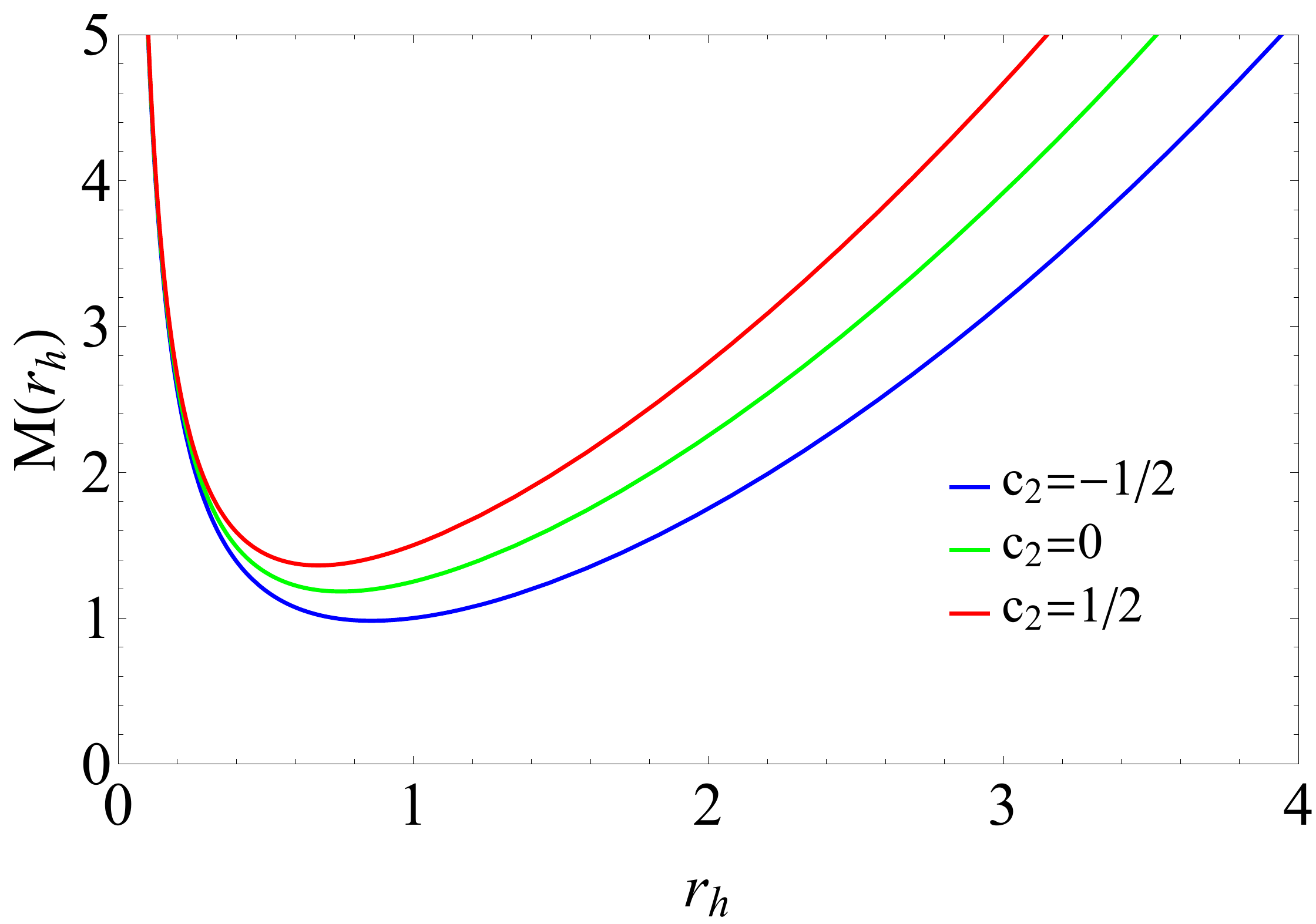}
\end{tabular}
 \caption{The black hole mass function $M(r_h)$ versus the event horizon radius $r_h$, which is inferred from the equation $f(r_h)=0$, at $Q=1$, $m_g=1$, and $c_1=1$.}\label{ehrad-mass-rela}
\end{figure}

From the behavior of the black hole mass function in terms of the horizon radius as depicted in Fig. \ref{ehrad-mass-rela}, we see that if the black hole mass is larger than a minimum value $M_{\text{min}}$, the black hole would possess two different horizons which are the event (outer) horizon $r_+$ and the Cauchy (inner) horizon $r_-$, corresponding to the non-extremal case. 
In this situation, by parameterizing the metric function $f(r)$ given in Eq. (\ref{fr-phir-exp}) in terms of the horizons $r_\pm$, $f(r)$ is rewritten as
\begin{eqnarray}
f(r)=\frac{m^2_gc_1}{2}\frac{(r-r_+)(r-r_-)(r+r_++r_-+\delta)}{r^2},\label{chbh-case}
\end{eqnarray}
where the ADM mass $M$ and the electric charge $Q$ of the black hole are expressed in terms of the horizons $r_\pm$ as
\begin{eqnarray}
M&=&\frac{m^2_gc_1}{4}\left[(r_++r_-)(r_++r_-+\delta)-r_+r_-\right],\nonumber\\
Q^2&=&\frac{m^2_gc_1}{2}r_+r_-(r_++r_-+\delta).
\end{eqnarray}
The tortoise coordinate is defined as
\begin{eqnarray}
r_*=\int\frac{dr}{f(r)}&=&\frac{1}{2\kappa_+}\log\frac{|r-r_+|}{r_+}+\frac{1}{2\kappa_-}\log\frac{|r-r_-|}{r_-}-\frac{m^2_gc_1}{8}\frac{(r_++r_-+\delta)^2(r_+-r_-)^2}{r^2_+r^2_-\kappa_+\kappa_-}\nonumber\\
&&\times\log\left(\frac{r+r_++r_-+\delta}{r_++r_-+\delta}\right),
\end{eqnarray}
where $\kappa_\pm$ are the surface gravity at the horizons $r_\pm$ and are given by
\begin{eqnarray}
\kappa_+&=&\frac{m^2_gc_1}{4r^2_+}(r_+-r_-)(2r_++r_-+\delta),\nonumber\\
\kappa_-&=&\frac{m^2_gc_1}{4r^2_-}(r_--r_+)(2r_-+r_++\delta).
\end{eqnarray}
In the Kruskal coordinate defined as
\begin{eqnarray}
U=-e^{-\kappa_+u}, \ \ \ \ V=e^{\kappa_+v},
\end{eqnarray}
the line element of the non-extremal charged black hole is rewritten as
\begin{eqnarray}
ds^2=-W^2(r)dUdV+r^2d\Omega^2_2,\label{Krcood-CBH}
\end{eqnarray}
where the conformal factor $W^2(r)$ reads
\begin{eqnarray}
W^2(r)&=&\frac{m^2_gc_1}{2}\frac{r_+r_-(r_++r_-+\delta)}{(\kappa_+r)^2}\left(\frac{r_-}{r-r_-}\right)^{\frac{\kappa_+}{\kappa_-}-1}\nonumber\\
&&\times\left(\frac{r_++r_-+\delta}{r+r_++r_-+\delta}\right)^{\frac{(r_++r_-+\delta)^2(r_+-r_-)}{r^2_+(2r_-+r_++\delta)}-1}.
\end{eqnarray}

Now we consider the case that the black hole mass is equal to the minimum value $M_{\text{min}}$, corresponding to the extremal case. In this situation, two horizons $r_\pm$ coincide together and hence the black hole possesses only one horizon at
\begin{eqnarray}
r_+=r_-\equiv r_e.
\end{eqnarray}
The metric function $f(r)$ in Eq. (\ref{chbh-case}) thus becomes
\begin{eqnarray}
f(r)=\frac{m^2_gc_1}{2}\frac{(r-r_e)^2(r+2r_e+\delta)}{r^2},
\end{eqnarray}
which corresponds to the tortoise coordinate as
\begin{eqnarray}
r_*&=&\frac{2}{m^2_gc_1}\left[-\frac{r^2_e}{(r-r_e)(3r_e+\delta)}+\frac{r_e(5r_e+2\delta)}{(3r_e+\delta)^2}\log\frac{|r-r_e|}{r_e}\right.\nonumber\\
&&\left.+\left(\frac{2r_e+\delta}{3r_e+\delta}\right)^2\log\left(\frac{r+2r_e+\delta}{2r_e+\delta}\right)\right].
\end{eqnarray}
Then, in terms of the Kruskal coordinate defined 
as
\begin{eqnarray}
U=-e^{-\frac{m^2_gc_1(3r_e+\delta)}{4r_e}u}, \ \ \ \ V=e^{\frac{m^2_gc_1(3r_e+\delta)}{4r_e}v},
\end{eqnarray}
we rewrite the line element of the extremal charged black hole as 
\begin{eqnarray}
ds^2=-W^2(r)dUdV+r^2d\Omega^2_2,\label{Krcood-ECBH}
\end{eqnarray}
where the conformal factor $W^2(r)$ is given by
\begin{eqnarray}
W^2(r)&=&\frac{8}{m^2_gc_1}\frac{r^4_e(2r_e+\delta)}{r^2(3r_e+\delta)^2}\left(\frac{r-r_e}{r_e}\right)^{\frac{r_e}{3r_e+\delta}}\left(\frac{r+2r_e+\delta}{2r_e+\delta}\right)^{1-\frac{(2r_e+\delta)^2}{r_e(3r_e+\delta)}}e^{\frac{r_e}{r-r_e}}.
\end{eqnarray}
\section{\label{EEforNBH} Entanglement entropy for neutral black hole}
In this section, we are interested in evaluating the entanglement entropy of Hawking radiation which is emitted by the black hole and identified as the matter sector coupled to gravity. We consider the contribution to the entanglement entropy from the configurations without and with the islands. For the case of the island configuration, we will restrict to the calculation in the presence of one island for simplification without loss of generality. We will indicate that the configuration of one island is sufficient to reproduce the finiteness for the entanglement entropy at the late times and thus it would lead to the Page curve consistent with the unitarity principle.

\begin{figure}[t]
 \centering
\begin{tabular}{cc}
\includegraphics[width=0.45 \textwidth]{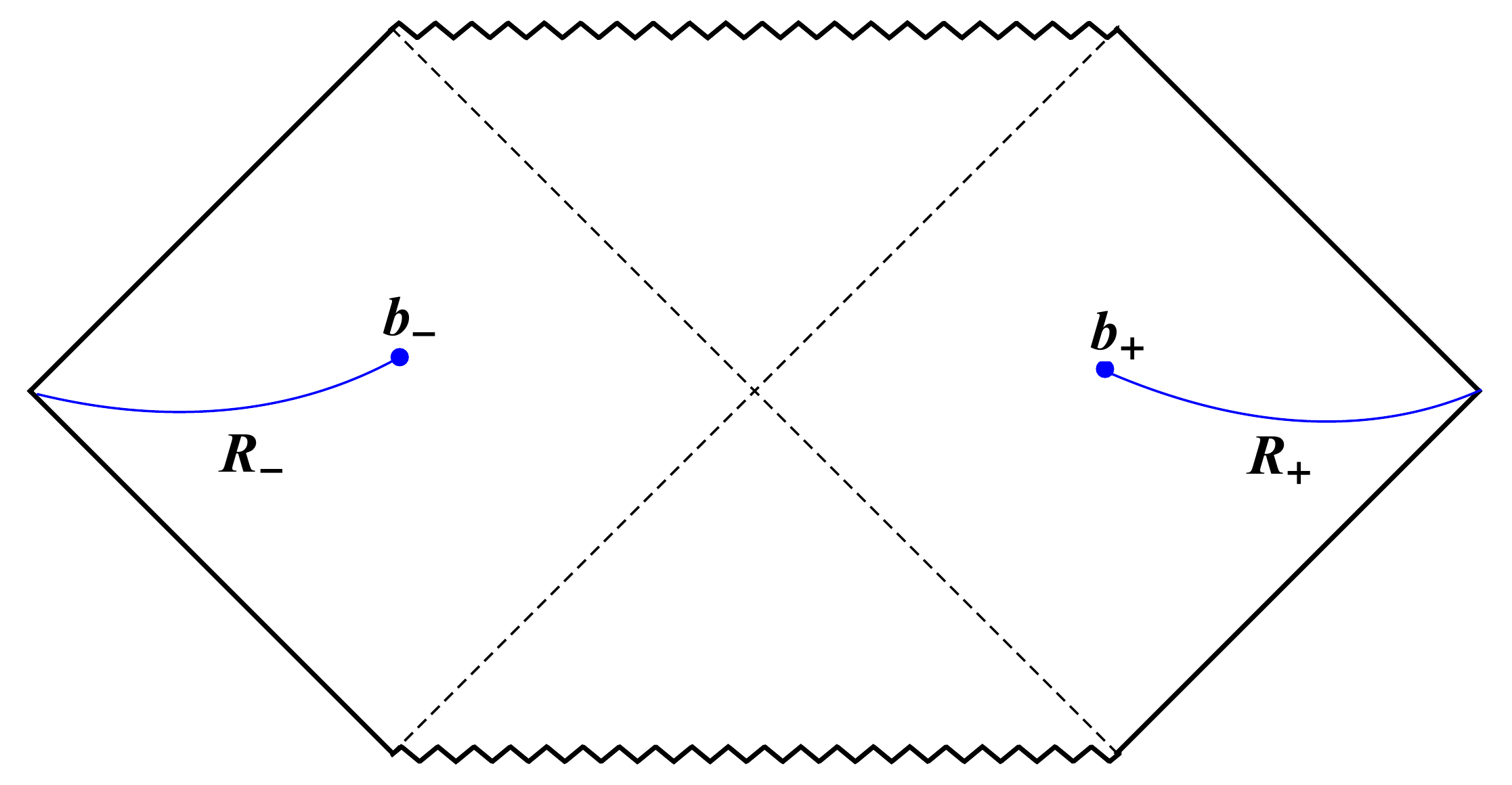}
\hspace*{0.05\textwidth}
\includegraphics[width=0.45 \textwidth]{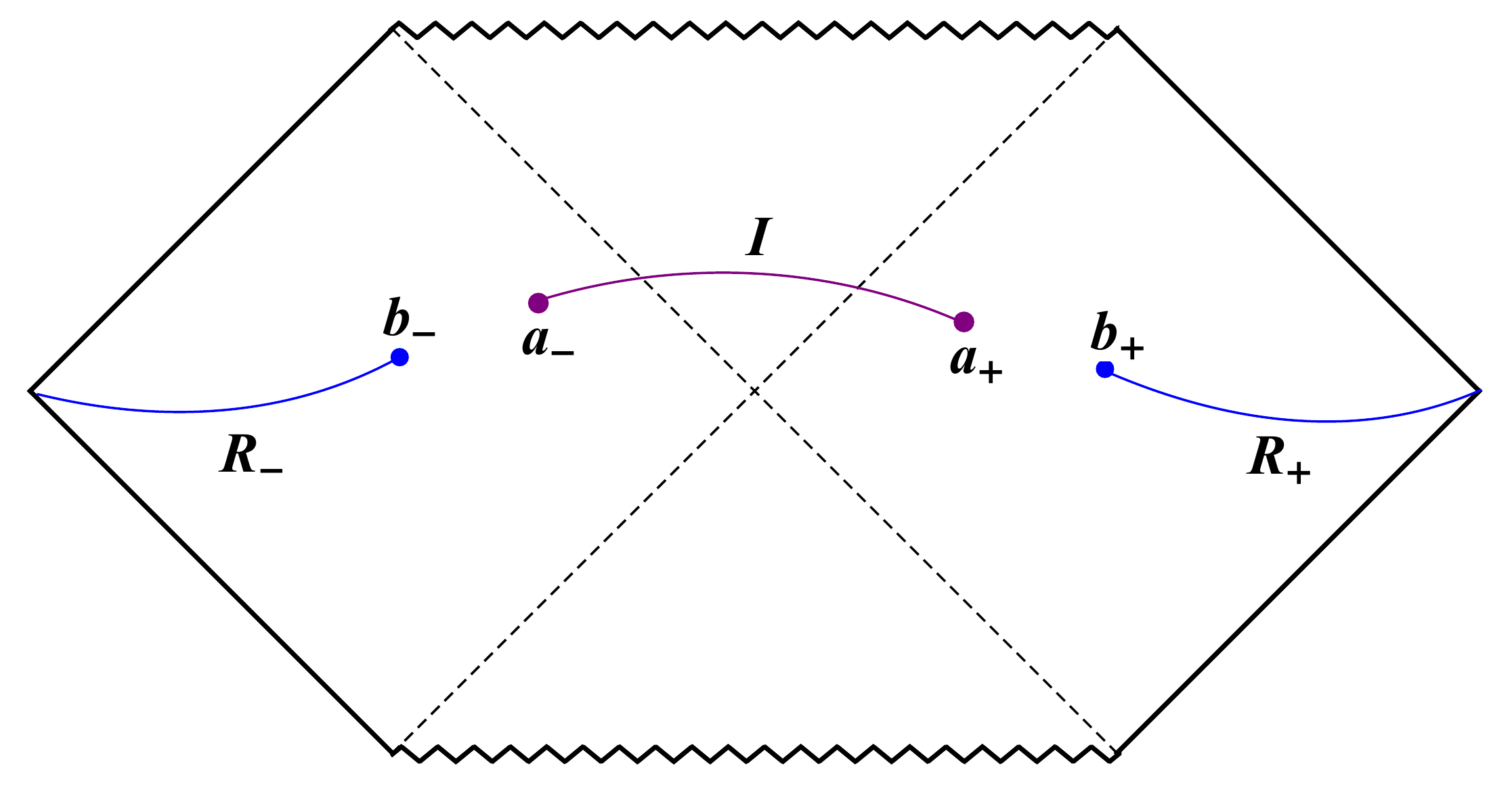}
\end{tabular}
  \caption{The Penrose diagram of the eternal neutral black hole in massive gravity in the absence (left panel) and the presence (right panel) of the island. The region of Hawking radiation consists of two parts $R_-$ and $R_+$ which lie on the left and right wedges, respectively. $b_-$ and $b_+$ represent the boundaries (cutoff surfaces) of $R_-$ and $R_+$, respectively. $I$ refers to the island, extending between the left and right wedges, whose boundaries are denoted by $a_\pm$.}\label{PD-neuBH}
\end{figure}

The Penrose diagram of the eternal neutral black hole (which is in the thermal equilibrium with a heat bath) without the island in massive gravity is depicted in the left panel of Fig. \ref{PD-neuBH} corresponding to the maximally extended spacetime $-\infty<(V+U)/2,(V-U)/2<+\infty$. 
The region of the Hawking radiation is given by the union of two regions $R_-$ and $R_+$ which are located in the left and right wedges of the Penrose diagram, respectively. The boundaries of the regions $R_-$ and $R_+$, which are introduced to avoid the IR divergence, are denoted by $b_-$ and $b_+$, respectively. The $(t,r)$ coordinates for $b_+$ and $b_-$ are $(t_b,b)$ and $(-t_b+i\beta/2,b)$, respectively, where $\beta$ is the inverse of the Hawking temperature of the black hole. For the region of Hawking radiation which is sufficiently far away from the black hole, we can ignore the backreaction of the matter sector on the spacetime geometry. With the distance between two boundaries of $b_+$ and $b_-$ to be large enough compared to the scale of the size of these boundaries, we can consider the approximation of the two-dimensional conformal field theory (CFT) to compute the entanglement entropy \cite{Hashimoto2020}. We also assume that the initial state of the system is in the pure state and hence the entanglement entropy in the radiation region $R_-\cup R_+$ is equal to that in its complement which is one interval $[b_-,b_+]$. In the absence of the island, the entanglement entropy is given by \cite{Almheiri2019,Fiola1994}
\begin{eqnarray}
S_{\text{mat}}(R_-\cup R_+)&=&\frac{c}{3}\log d(b_-,b_+)\nonumber\\
&=&\frac{c}{12}\log\left[W^2(b_-)W^2(b_+)(U(b_-)-U(b_+))^2(V(b_-)-V(b_+))^2\right],\label{ent-noland}
\end{eqnarray}
where $c$ is the central charge of the two-dimensional CFT and $d(b_-,b_+)$ is the geodesic distance between $b_-$ and $b_+$ which is calculated for the metric of the following form $ds^2=-W^2dUdV$.

In the presence of an island, the corresponding Penrose diagram is shown in the right panel of Fig. \ref{PD-neuBH}.
The boundaries of the island located in the left and right wedges of the Penrose diagram are represented by $a_+$ and $a_-$ whose $(t,r)$ coordinates are $(t_a,a)$ and $(-t_a+i\beta/2,a)$, respectively. In the island construction, the generalized entropy is given as 
\begin{eqnarray}
S_{\text{gen}}=\frac{2\pi a^2}{G_N}+S_{\text{mat}}(R_-\cup R_+\cup I),\label{gen-ent-island}
\end{eqnarray}
where the first term is the Bekenstein-Hawking entropy coming from the contribution of two island boundaries and the entropy of the matter sector $S_{\text{mat}}(R_-\cup R_+\cup I)$ is calculated by the formula for two intervals as \cite{Casini2005}
\begin{eqnarray}
S_{\text{mat}}(R_-\cup R_+\cup I)=\frac{c}{3}\log\left[\frac{d(a_+,a_-)d(b_+,b_-)d(a_+,b_+)d(a_-,b_-)}{d(a_+,b_-)d(a_-,b_+)}\right].\label{Sma-isl}
\end{eqnarray}
According to the prescription of the quantum extremal surface, the dominant contribution for the entanglement entropy comes from the configuration which minimizes the generalized entropy. Therefore, we shall compute the entanglement entropy by extremizing the generalized entropy over all possible boundary surfaces of the island and then take the minimal value.

\subsection{Entanglement entropy without island}

First we study the behavior of the entanglement entropy of Hawking radiation in the case of the no island configuration. From Eq. (\ref{ent-noland}), one obtains an explicit expression of the entanglement entropy for the neutral black hole in massive gravity in the absence of the island as
\begin{eqnarray}
S_{\text{mat}}(R_-\cup R_+)&=&\frac{c}{6}\log\left[\frac{4f(b)}{\kappa^2}\cosh^2\kappa t_b\right]\nonumber\\
&=&\frac{c}{6}\log\left[\frac{32}{m^2_gc_1}\frac{r^2_h(b-r_h)(b+r_h+\delta)}{b(2r_h+\delta)^2}\cosh^2\kappa t_b\right]\nonumber\\
&\simeq&\frac{c}{6}\log\left[\frac{32}{m^2_gc_1}\frac{r^2_h(b+\delta)}{(2r_h+\delta)^2}\cosh^2\kappa t_b\right],
\end{eqnarray}
where we have used $r_h\ll b$ in third line due to the boundaries of the radiation region assumed to be far away from the black hole. At the early time approximation $t_b\ll1/\kappa$ or $t_b\ll r_h$, we have
\begin{eqnarray}
S_{\text{mat}}(R_-\cup R_+)&\simeq&\frac{c}{6}\log\left[\frac{32}{m^2_gc_1}\frac{r^2_h(b+\delta)}{(2r_h+\delta)^2}\right]+\frac{c}{6}\left[\frac{m^2_gc_1(2r_h+\delta)}{4r_h}t_b\right]^2.\label{Ent-noland-et}
\end{eqnarray}
Here, the first term is the entanglement entropy of the radiation region at the initial state up to a constant and the second term manifests the quadratic growth of the entanglement entropy with time at the early times. For $t_b\gg1/\kappa$ corresponding to the late times, we find
\begin{eqnarray}
S_{\text{mat}}(R_-\cup R_+)&\simeq&\frac{m^2_gc_1(2r_h+\delta)c}{12r_h}t_b.\label{NBHent-early}
\end{eqnarray}
This expression implies that the entanglement entropy of Hawking radiation increases linearly in time and becomes infinite as $t_b\rightarrow\infty$, which corresponds to the fact that the distance between two regions $R_-$ and $R_+$ is very large at the late times. The infinitely large value of the entanglement entropy in the limit $t_b\rightarrow\infty$  means that the information about the initial-state matter which collapses into the black hole or the information about the particles falling into the black hole cannot be retrieved from the Hawking radiation. In this sense, the contribution of the no island configuration to the entanglement entropy leads to the non-unitary time evolution of the black hole in the evaporation process.
In the next subsection, we will show that this conflicting issue with the unitarity of quantum mechanics can be resolved with the island configuration which emerges at the late times of the evaporation process.

\subsection{Entanglement entropy with an island}

We calculate the entanglement entropy of Hawking radiation with including the contribution of the configuration with an island. From Eqs. (\ref{gen-ent-island}) and (\ref{Sma-isl}), we can write the generalized entropy in the presence of one island for the metric (\ref{Krcood-NBH}) as
\begin{eqnarray}
S_{\text{gen}}&=&\frac{2\pi a^2}{G_N}+\frac{c}{6}\log\left[16W^2(a)W^2(b)e^{2\kappa(r_*(a)+r_*(b))}\cosh^2(\kappa t_a)\cosh^2(\kappa t_b)\right]\nonumber\\
&&+\frac{c}{3}\log\left[\frac{\cosh(\kappa(r_*(a)-r_*(b)))-\cosh(\kappa(t_a-t_b))}{\cosh(\kappa(r_*(a)-r_*(b)))+\cosh(\kappa(t_a+t_b))}\right],\label{Sgen-isl-NBH}
\end{eqnarray} 
where the first term comes from the two-sided area of the boundaries of the island and the second and third terms are the contributions of the matter fields on the union of the radiation and island regions.

In the presence of the island, let us study the behavior of the entanglement entropy at the early and late times of the black hole evaporation:

\textbf{Early times} At the early times of the black hole evaporation, the entanglement entropy of Hawking radiation is small and hence the island should be lie inside the black hole. In addition, we assumed the boundaries of the radiation region far away from the event horizon of the black hole. These lead to the following approximation \cite{Hashimoto2020,LiWang2021}
\begin{eqnarray}
r_h\ll b, \ \ \ \ t_a,t_b\ll1/\kappa\ll r_*(b)-r_*(a).
\end{eqnarray}
As a result, we can properly neglect the third term in the expression (\ref{Sgen-isl-NBH}) and thus we obtain
\begin{eqnarray}
S_{\text{gen}}&=&\frac{2\pi a^2}{G_N}+\frac{c}{3}\left[\kappa r_*(a)+\log W(a)\right]+\frac{c}{6}(\kappa t_a)^2+\cdots\nonumber\\
&\simeq&\frac{2\pi a^2}{G_N}+\frac{c}{6}\left[\log|a-r_h|+\log(r_h+a+\delta)-\log a\right]+\frac{c}{6}(\kappa t_a)^2+\cdots\label{genentr-earlytime},
\end{eqnarray}
where the ellipses refer to the terms which are independent on $a$ and $t_a$. In order to find the position of the island boundaries which extremize the generalized entropy $S_{\text{gen}}$, we need to extremize $S_{\text{gen}}$ given in Eq. (\ref{genentr-earlytime}) over all possible $(a,t_a)$ locations of island according to the island method \cite{Engelhardt2015,Ryu2006,Hubeny2007,Akers2020,Faulkner2013,Wall2014}, which means that we need to solve the following extremizing equations 
\begin{eqnarray}
\frac{\partial S_{\text{gen}}}{\partial a}=0, \ \ \ \ \frac{\partial S_{\text{gen}}}{\partial t_a}=0.
\end{eqnarray}
Note that, on the black hole interior the radial coordinate $r$ is actually the timelike coordinate. Hence, at the early times we have $a/r_h\ll1$. Using this approximation and $cG_N/r^2_h\ll1$, we obtain the location of the island boundaries as
\begin{eqnarray}
a\simeq\sqrt{\frac{c}{24\pi}}\left[1+\sqrt{\frac{cG_N}{6\pi r^2_h}}\frac{\delta}{4(r_h+\delta)}\right]l_P,\label{isl-si-et}
\end{eqnarray}
where $l_P\equiv\sqrt{G_N}$ is the Planck length. This result indicates that the size of the island at the early times is about the Planck length. However, the upper cutoff length in the derivation of the island formula should be far above the Planck length where the Planck scale physical degrees of freedom are integrated out. This can be realized from the assumption of the replica symmetry, which is broken by the effects of quantum gravity, to derive the island formula for the entanglement entropy \cite{Hartman2020,Stanford2019}. In this sense, the island would not emerge at the early times. Thus, the entanglement entropy at the early times should be determined by the geometry configuration without island and since it grows with time, as analyzed in the previous subsection. 

Furthermore, we note that massive gravity does not make a significant effect on the size of the island given in Eq. (\ref{isl-si-et}). This is realized from the fact that the nonzero mass of graviton can be considered as the IR or long-length corrections which hence do not affect significantly on the physics in the Planck scale region.

\textbf{Late times} We turn to consider the late time behavior of the entanglement entropy. In this situation, the approximation is taken as \cite{Hashimoto2020,LiWang2021}
\begin{eqnarray}
1/\kappa \ll r_*(b)-r_*(a)\ll t_a,t_b,\label{lat-isl-app}
\end{eqnarray}
which leads to
\begin{eqnarray}
\cosh\kappa t_{a,b}\simeq\frac{1}{2}e^{\kappa t_{a,b}},\ \ \ \ \cosh\kappa(t_a+t_b)\gg\cosh(\kappa(r_*(b)-r_*(a)).
\end{eqnarray}
With this approximation, the time-dependent component of the generalized entropy is approximated as
\begin{eqnarray}
S_{\text{t-dep}}\simeq\frac{c}{3}\log\left[\frac{\cosh(\kappa(r_*(a)-r_*(b)))-\cosh(\kappa(t_a-t_b))}{2}\right].
\end{eqnarray} 
From this expression, it is straightforward to find that extremizing the generalized entropy with respect to $t_a$ leads to $t_a=t_b$. Substituting this result into the generalized entropy within the approximation (\ref{lat-isl-app}), we obtain the following approximate expression
\begin{eqnarray}
S_{\text{gen}}&\simeq&\frac{2\pi a^2}{G_N}+\frac{c}{6}\log\left[W^2(a)W^2(b)\right]+\frac{2c}{3}\kappa r_*(b)-\frac{2c}{3}e^{-\kappa(r_*(b)-r_*(a))}\nonumber\\
&\simeq&\frac{2\pi a^2}{G_N}+\frac{c}{3}\log\left[\frac{8r^2_h(r_h+\delta)}{m^2_gc_1(2r_h+\delta)^2}\frac{(b-r_h)}{\sqrt{ab}}\left(\frac{a+r_h+\delta}{r_h+\delta}\right)^{-\frac{\delta}{2r_h}}\left(\frac{b+r_h+\delta}{r_h+\delta}\right)^{\frac{\delta+2r_h}{2r_h}}\right]\nonumber\\
&&-\frac{2c}{3}\left|\frac{a-r_h}{b-r_h}\right|^{1/2}\left(\frac{a+r_h+\delta}{b+r_h+\delta}\right)^{\frac{\delta+r_h}{2r_h}}.\label{NBH-genent-lt}
\end{eqnarray}
This expression suggests that the entanglement entropy is no longer dependent on time at the late times. 

We consider the situation that the island is located near the event horizon, i.e. $a=r_h+\epsilon+\mathcal{O}(\epsilon^2)$ with $\epsilon\ll1$. By solving perturbatively the extremizing condition $\partial S_{\text{gen}}/\partial a=0$, we find
\begin{eqnarray}
a\simeq r_h\left[1+\left(\frac{cG_N}{r^2_h}\right)^2\frac{1}{144\pi^2}\frac{r_h}{b-r_h}\left(\frac{2r_h+\delta}{b+r_h+\delta}\right)^{\frac{r_h+\delta}{r_h}}\right].\label{sil-loc-lt}
\end{eqnarray}
Clearly, the second term is positive and is suppressed by the second power of $cG_N/r^2_h$, which is really small as expected. This implies that the location of the island boundaries is slightly outside the event horizon of the black hole. We also observe that the nonzero mass of graviton affects directly and indirectly on the the location of the island boundaries through the terms relating to $\delta$ and the modification of the event horizon radius $r_h$, respectively. However, due to the fact that the second term in Eq. (\ref{sil-loc-lt}) is small, the direct effects of massive gravity on the location of the island boundaries are negligible compared to its indirect effects. From the expansion of the event horizon radius in terms of the graviton mass $m_g$ for the small $m_g$ situation (which is consistent with the fact) as
\begin{eqnarray}
r_h=2M\left[1-(c_2+c_1M)m^2_g+\mathcal{O}(m^4_g)\right],\label{ep-evh}
\end{eqnarray}
we realize that the nonzero mass of graviton which makes the location of the island boundaries shifting inside or outside the black hole depends on the sign of the term $c_2+c_1M$. If the black hole mass is larger than the ratio $-c_2/c_1$ or the event horizon radius of the corresponding Schwarzschild black hole (in Einstein gravity) is larger than the ratio $-2c_2/c_1$, the location of the island boundaries would be closer to the event horizon of the black hole compared to Einstein gravity. On the contrary, the location of the island boundaries would be pushed further.

In addition, as the location of the cutoff surface approaches the event horizon, the second term in Eq. (\ref{sil-loc-lt}) becomes large and hence it is no longer considered as the correction. This means that in this situation the boundaries of the island are not close to the event horizon. Consequently, some part of the island would lie inside the region of Hawking radiation and hence the concept of the island here does not make sense. On the other hand, the calculation of the entanglement entropy using the island formula is invalid if the radiation region is close to the event horizon.

By substituting the location of the island boundaries given in Eq. (\ref{sil-loc-lt}) into the approximate expression of the generalized entropy at the late times given in Eq. (\ref{NBH-genent-lt}), we determine the entanglement entropy as
\begin{eqnarray}
S_{EE}=\frac{2\pi r^2_h}{G_N}+\frac{c}{3}\log\left[\frac{8r^{3/2}_h(b-r_h)}{m^2_gc_1(r_h+\delta)\sqrt{b}}\left(\frac{2r_h+\delta}{r_h+\delta}\right)^{-\frac{\delta+4r_h}{2r_h}}\left(\frac{b+r_h+\delta}{r_h+\delta}\right)^{\frac{\delta+2r_h}{2r_h}}\right]+\mathcal{O}\left(\frac{cG_N}{r^2_h}\right)\label{NBHent-late}.\nonumber\\
\end{eqnarray}
Here, the first term is twice of the Bekenstein-Hawking entropy of the black hole, which comes from the area of two-sided island and it is the dominant contribution to the entanglement entropy. On the other hand, the entanglement entropy which is observed in a single side of the Penrose diagram is approximately the Bekenstein-Hawking entropy. The second term is the logarithm correction for the entanglement entropy, which comes from the quantum nature of the matter fields. Other correction terms are strongly suppressed by the powers of $cG_N/r^2_h$ and thus they are very small. In this way, the presence of the island which becomes the dominant configuration at the late times leads a finite saturation value for the entanglement entropy which is approximated at the leading order to be twice of the Bekenstein-Hawking entropy. The expansion (\ref{ep-evh}) implies that at the leading order the nonzero mass of graviton reduces(raises) the bound of the entanglement entropy of Hawking radiation compared to the massless case if the black hole mass is larger(smaller) than the ratio $-c_2/c_1$.

In summary, we have calculated the entanglement entropy of Hawking radiation by extremizing the generalized entropy over all possible boundary surfaces of the island from which we determine the location of the island boundaries and then we obtain a minimum value. At the early stage of the black hole evaporation, no island is emerged and the configuration without the island minimizes the generalized entropy. As a result, in during this moment the entanglement entropy is approximately a linearly increasing function in time. However, this behavior of the entanglement entropy changes drastically at the late stage of the black hole evaporation as the island appears with its boundaries located slightly outside the event horizon. The configuration with the island leads to a minimum value of the generalized entropy which is twice of the Bekenstein-Hawking entropy of the black hole at the leading order. Interestingly, transition between the linear growth and time-independent constant behaviors of the entanglement entropy can be realized from Page's argument that the entanglement entropy can be approximately given by the thermal entropy of the subsystem if it is sufficiently small compared to the total system \cite{Page1993a,Page1993b}. In the beginning of the black hole evaporation where the amount of Hawking radiation emitted by the black hole and entering into the boundaries of the radiation region is still small, the radiation region is substantially smaller than the total system. Thus, the entanglement entropy can be approximated by the thermal entropy of the radiation which grows with time due to the increasing of the amount of the radiation. But, as the amount of the radiation becomes more and more at the late stage of the black hole evaporation, the subsystem would be replaced by the black hole and hence the entanglement entropy stops the growth with time and reaches the bound of the Bekenstein-Hawking entropy of the black hole. 

We can check that our calculations for the location of the island boundaries and the entanglement entropy with the island configuration would reduce to the corresponding results of the four-dimensional Schwarzschild black hole \cite{Hashimoto2020} in the limit of that the graviton mass goes to zero, i.e. $m_g\rightarrow0$. By using the fact that $m_g\rightarrow0$ corresponds to $\delta\rightarrow\infty$ and the following limit 
\begin{eqnarray}
\lim_{\delta\rightarrow\infty}\left(1+\frac{x}{\delta}\right)^{\delta}=e^x,
\end{eqnarray}
we find the results in the case of the four-dimensional Schwarzschild black hole as follows
\begin{eqnarray}
a&\longrightarrow& r_h+\frac{(cG_N)^2}{144\pi^2 r^2_h(b-r_h)}e^{\frac{r_h-b}{r_h}},\nonumber\\
S_{EE}&\longrightarrow&\frac{2\pi r^2_h}{G_N}+\frac{c}{6}\log\left[\frac{16r^3_h(b-r_h)^2}{b}e^{\frac{b-r_h}{r_h}}\right]+\mathcal{O}\left(\frac{cG_N}{r^2_h}\right),
\end{eqnarray}
where $r_h=2M$.

\subsection{Page time and scrambling time}

\begin{figure}[t]
 \centering
\begin{tabular}{cc}
\includegraphics[width=0.5 \textwidth]{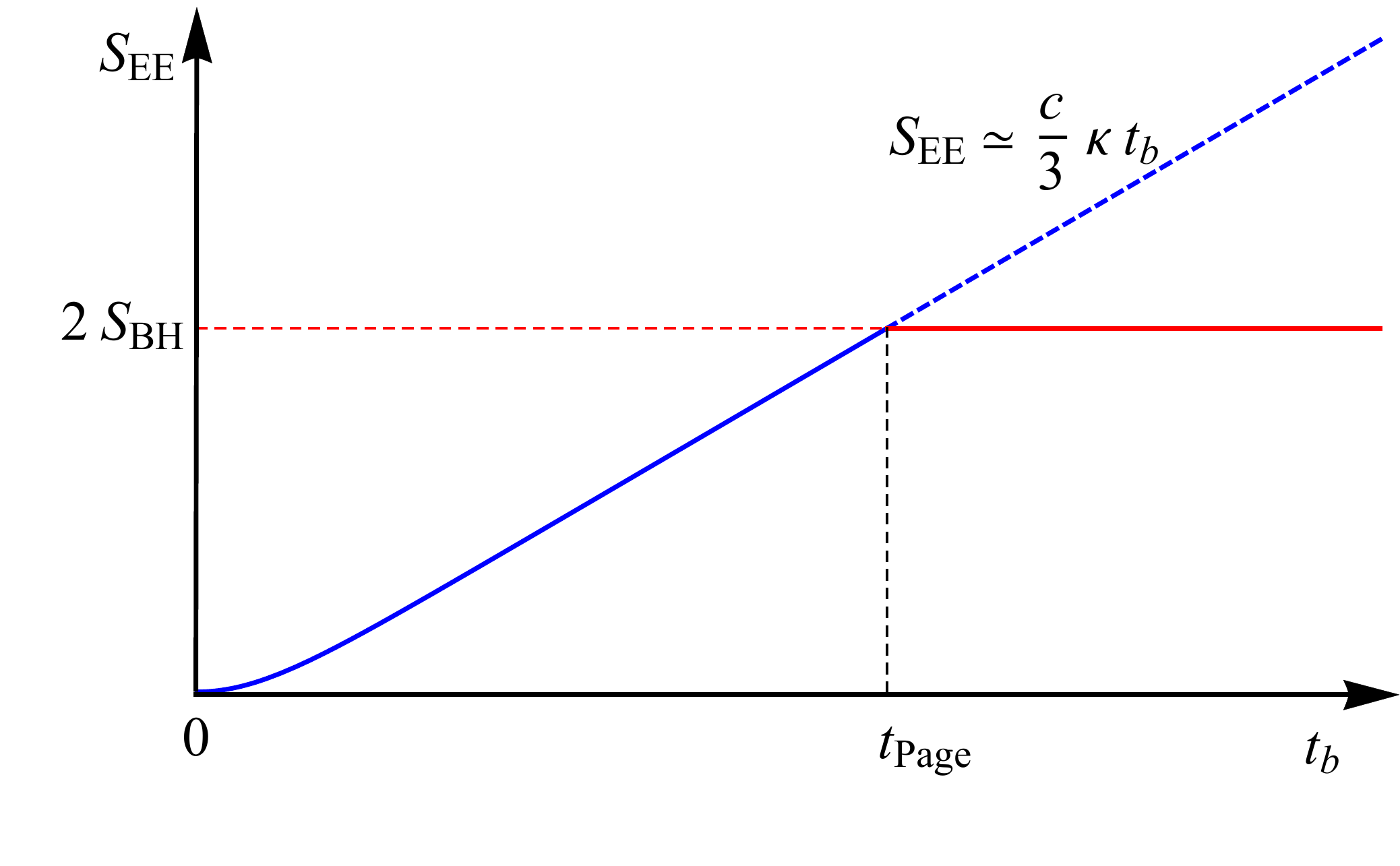}
\end{tabular}
 \caption{The Page curve for the neutral black hole in massive gravity. The solid and dashed blue lines refer to the time evolution of the entanglement entropy at the early times and the late times without the island, respectively. The solid red line stands for the entanglement entropy at the late times in the presence of an island.}\label{Page-curve}
\end{figure}

The results of the previous subsections allow us to sketch the behavior of the entanglement entropy in time as shown in Fig. \ref{Page-curve}. 
We observe that, at the first stage of the black hole evaporation, the entanglement entropy grows linearly with time due to the dominant contribution of the configuration without the island. At the late stage of the black hole evaporation, the island emerges near the event horizon and becomes the preferred configuration. As a result, the linear growth of the entanglement entropy reaches maximum and is replaced by a constant. 

The Page time is the moment that the entanglement entropy reaches the maximal value corresponding to the transition between two configurations without and with the island. From Fig. \ref{Page-curve}, we see that the Page time can be approximately calculated from the crossing of the solid blue line (the early times without the island) and the solid red line (the late times with the island). On the other hand, by equating Eqs. (\ref{NBHent-early}) and (\ref{NBHent-late}), we derive the Page time as
\begin{eqnarray}
t_{\text{Page}}\simeq\frac{24\pi r^3_h}{m^2_gc_1(2r_h+\delta)cG_N}=\frac{3S_{\text{BH}}}{\pi c T_H},
\end{eqnarray}
where $T_H=\kappa/2\pi$ is the Hawking temperature of the neutral black hole. In order to see the effect of the graviton mass $m_g$ on the Page time, let us expand the expression of the Page time just obtained in terms of $m_g$ as
\begin{eqnarray}
t_{\text{Page}}=\frac{96\pi M^3}{cG_N}\left[1-(4c_2+5c_1M)m^2_g+\mathcal{O}(m^4_g)\right],\label{Exp-Pati-NBH}
\end{eqnarray}
where the first term is the Page time for the Schwarzschild black hole in the context of Einstein gravity (corresponding to the massless graviton) and the second and higher order terms are the contributions due to the nonzero mass of graviton. The expression (\ref{Exp-Pati-NBH}) suggests that the nonzero mass of graviton makes the evaporation of the neutral black hole more quickly to reach the Page time compared to the massless case if the black hole mass $M$ is larger than the ratio $-4c_2/5c_1$. Of course, this happens the contrary for $M<-4c_2/5c_1$. The reduction (or the increasing) of the Page time in massive gravity with the proper coupling parameters $c_{1,2}$ can be understood as follows. First, with $M>-c_2/c_1$ the Bekenstein-Hawking entropy of the black hole or the bound of the entanglement entropy in massive gravity is smaller than that in Einstein gravity. This means that the system in massive gravity takes a shorter duration to reach the entropy bound if the Hawking temperature of the black hole in two theories of gravity is the same. Second, by expanding the Hawking temperature of the black hole in massive gravity as
\begin{eqnarray}
T_H=\frac{1}{8M\pi}\left[1+(2c_2+3c_1M)m^2_g+\mathcal{O}(m^4_g)\right],
\end{eqnarray}
where the first term is the Hawking temperature of the conventional Schwarzschild black hole and the remaining terms come from the presence of the graviton mass, we find that the Hawking temperature of the neutral black hole in massive gravity is larger than that in Einstein gravity in the region of $M>-2c_2/3c_1$. The larger temperature means that the black hole emits the Hawking radiation more rapidly and hence it yields the appearance of the island more early. Combining the behavior of both the Bekenstein-Hawking entropy and the Hawking temperature in terms of the coupling parameters of massive gravity leads the intermediate value $-4c_2/5c_1$ which lies between $-c_2/c_1$ and $2c_2/3c_1$.

The presence of the island reproduces the behavior of the entanglement entropy following the Page curve. Thus, we can consider the scrambling time which is defined as the minimum time for the recovery of the information which can be retrieved from the Hawking radiation after falling into the black hole according to the Hayden-Preskill protocol \cite{Hayden2007}. Recall that, in the entanglement wedge construction according to the prescription of the island, the density matrix of the Hawking radiation is represented by the union of the radiation region $R_-\cup R_+$ and the island. This implies that the information of the signal which is thrown into the island is not contained in the black hole but it could be decoded by the Hawking radiation. If an observer sends a signal from the cutoff surface, it would reach the island after an earliest duration $t_{\text{scr}}$ identified as the scrambling time defined as
\begin{eqnarray}
t_{\text{scr}}=r_*(b)-r_*(a)&=&\frac{1}{2\kappa}\left[\log\left|\frac{b-r_h}{a-r_h}\right|+\left(1+\frac{\delta}{r_h}\right)\log\left(\frac{b+r_h+\delta}{a+r_h+\delta}\right)\right]\nonumber\\
&\simeq&\frac{1}{\kappa}\log\frac{r^2_h}{G_N}\simeq\frac{1}{2\pi T_H}\log S_{\text{BH}}.
\end{eqnarray}
The leading order contribution for the scrambling time is proportional to the inverse of the Hawking temperature and the logarithm of the black hole entropy, which is consistent with the result obtained in Ref. \cite{Sekino2008}. In this way, the expression of the scrambling time at the leading order is universal. Furthermore, we observe that the scrambling time is very small compared to the Page time. In order to see the explicit influence of the nonzero mass of graviton on the scrambling time, we expand the expression for $t_{\text{scr}}$ as
\begin{eqnarray}
t_{\text{scr}}=4M\log\frac{4\pi M^2}{G_N}-4M\left[2(c_2+c_1M)+(2c_2+3c_1M)\log\frac{4\pi M^2}{G_N}\right]m^2_g+\mathcal{O}(m^4_g),
\end{eqnarray}
where the first term is the prediction of Einstein gravity. This expression implies that the presence of the graviton mass would reduce the scrambling time if the ratio of the massive gravity coupling parameters $c_{1,2}$ satisfies
\begin{eqnarray}
\frac{c_2}{c_1}>-\frac{M}{2}\left(3-\frac{1}{1+\log\frac{4\pi M^2}{G_N}}\right).
\end{eqnarray}
\section{\label{EEforCBH} Entanglement entropy for charged black holes}
In this section, we shall compute the entanglement entropy of Hawking radiation with the absence and presence of the islands, the Page time, and the scrambling time for the non-extremal and extremal charged black holes in massive gravity. The arguments and the method which are presented in the case of the neutral black hole can still be applied for the situation of the nonzero electric charge. There are some differences for the extremal charged black hole, which are due to its Penrose diagram to be a one-sided geometry, rather than the two-sided geometry like the Penrose diagram of the neutral and non-extremal charged black holes.

\subsection{Non-extremal case}
\begin{figure}[t]
 \centering
\begin{tabular}{cc}
\includegraphics[width=0.45 \textwidth]{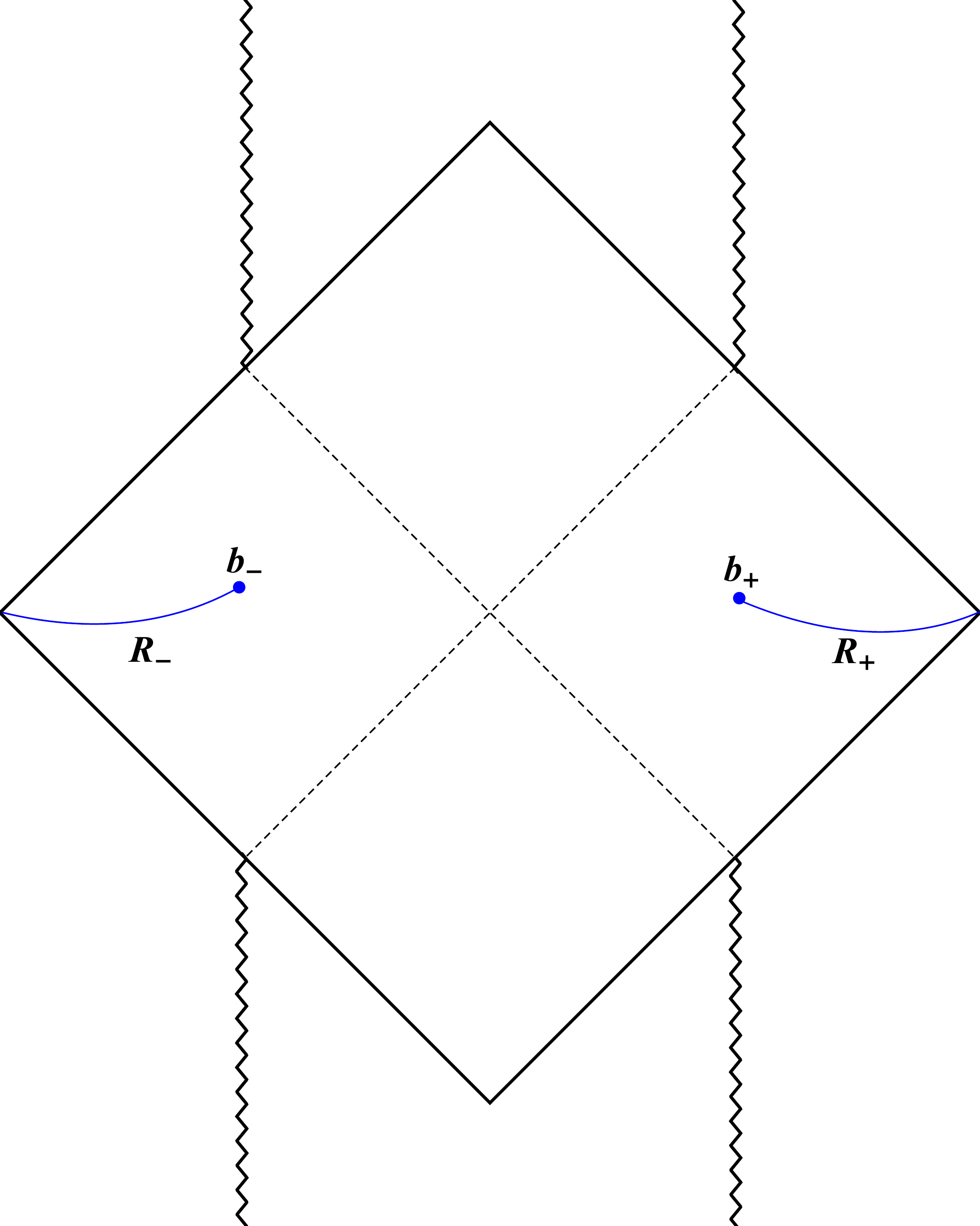}
\hspace*{0.05\textwidth}
\includegraphics[width=0.45 \textwidth]{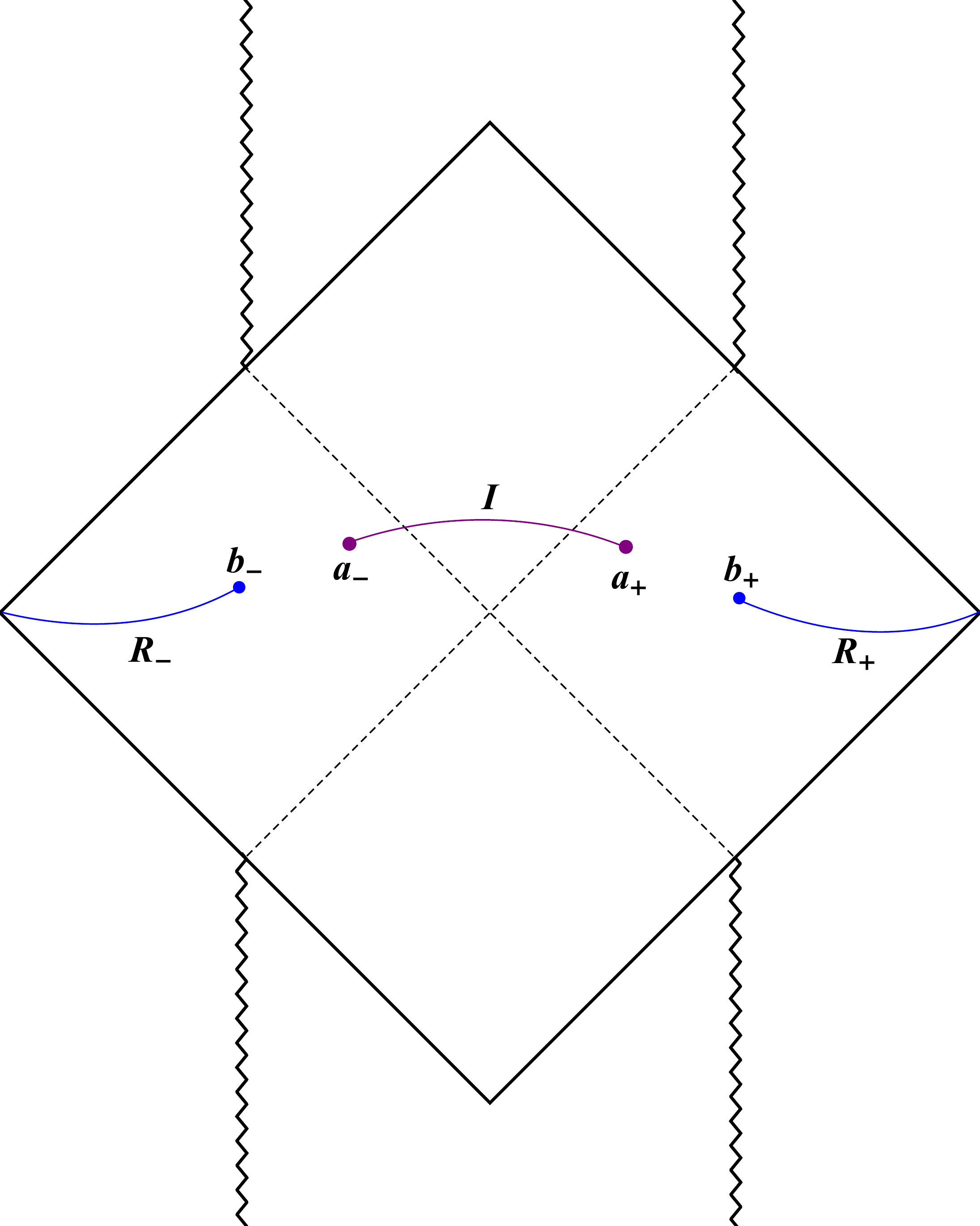}
\end{tabular}
  \caption{The Penrose diagram of the eternal non-extremal charged black hole in massive gravity with the no island configuration (left panel) and the island configuration (right panel). The radiation region is the union of two parts $R_\pm$ whose boundaries are denoted by $b_\pm$. $I$ refers to the island with the boundaries represented by $a_\pm$.}\label{Pd-ne-CBH}
\end{figure}

First, we consider the non-extremal charged black hole whose metric in the Kruskal coordinate is given by (\ref{Krcood-CBH}). The Penrose diagram without and with the island configuration is depicted in the left and right panels of Fig. \ref{Pd-ne-CBH}, respectively.

For the case of the configuration with no island, the entanglement entropy is obtained as
\begin{eqnarray}
S_{\text{mat}}(R_-\cup R_+)&=&\frac{c}{6}\log\left[\frac{4f(b)}{\kappa^2_+}\cosh^2\kappa_+t_b\right]\nonumber\\
&\simeq&\frac{c}{6}\log\left[\frac{32}{m^2_gc_1}\frac{r^4_+(b+\delta)}{(r_+-r_-)^2(2r_++r_-+\delta)^2}\cosh^2\kappa_+t_b\right],
\end{eqnarray}
where we have used the approximation $r_h\ll b$. In analogy to the neutral black hole which we study in the previous section, the entanglement entropy for the non-extremal charged black hole with no island grows linearly with time in the limit of $1/\kappa_+\ll t_b$ and thus it becomes infinite at the late stage of the evaporation. Of course, this is inconsistent with the unitarity time evolution. Therefore, we expect the emergence of the island configuration which minimizes the generalized entropy by which the entanglement entropy stops the linear increasing and reaches a saturation value.

Now we arrive at the case of the configuration with one island. The generalized entropy with an island at the early times reads
\begin{eqnarray}
S^{\text{et}}_{\text{gen}}&=&\frac{2\pi a^2}{G_N}+\frac{c}{3}\left[\kappa_+r_*(a)+\log W(a)\right]+\frac{c}{6}(\kappa_+t_a)^2+\cdots\nonumber\\
&\simeq&\frac{2\pi a^2}{G_N}+\frac{c}{6}\left[\log|a-r_+|+\log|a-r_-|+\log(a+r_++r_-+\delta)-2\log a\right]\nonumber\\
&&+\frac{c}{6}(\kappa_+t_a)^2+\cdots,
\end{eqnarray}
where the ellipses refer to the terms independent on the temporal and spatial location of the island boundaries. By extremizing $S^{\text{et}}_{\text{gen}}$ with respect to $a$, we find
\begin{eqnarray}
a\simeq\sqrt{\frac{c}{12\pi}}\left[1+\sqrt{\frac{cG_N}{3\pi r^2_+}}\frac{r^2_++r^2_-+r_+r_-+\delta(r_++r_-)}{8r_-(r_++r_-+\delta)}\right]l_P,
\end{eqnarray}
which implies that the size of the island in the beginning of the black hole evaporation is in order of the Planck length. On the other hand, no island emerges at the early times and thus the behavior of the entanglement entropy is governed by the configuration with no island.

Next we study the late time behavior of the entanglement entropy in the presence of the island. The generalized entropy in this situation is found as
\begin{eqnarray}
S^{\text{lt}}_{\text{gen}}&\simeq&\frac{2\pi a^2}{G_N}+\frac{c}{6}\log\left[W^2(a)W^2(b)\right]+\frac{2c}{3}\kappa_+r_*(b)-\frac{2c}{3}e^{-\kappa_+(r_*(b)-r_*(a))}\nonumber\\
&\simeq&\frac{2\pi a^2}{G_N}+\frac{c}{6}\log\left[\frac{f(a)f(b)}{\kappa^4_+}e^{2\kappa_+(r_*(b)-r_*(a))}\right]-\frac{2c}{3}e^{-\kappa_+(r_*(b)-r_*(a))}\nonumber\\
&\simeq&\frac{2\pi a^2}{G_N}+\frac{c}{3}\log\left[\frac{m^2_gc_1}{2}\frac{(b-r_+)(b-r_-)(b+r_++r_-+\delta)}{\kappa^2_+ab}\right]+\frac{c}{6}\left(\frac{\kappa_+}{\kappa_-}-1\right)\log\left|\frac{b-r_-}{a-r_-}\right|\nonumber\\
&&+\frac{c}{6}\left[\frac{(r_++r_-+\delta)^2(r_+-r_-)}{r^2_+(2r_-+r_++\delta)}-1\right]\log\left(\frac{b+r_++r_-+\delta}{a+r_++r_-+\delta}\right)\nonumber\\
&&-\frac{2c}{3}\left|\frac{a-r_+}{b-r_+}\right|^{\frac{1}{2}}\left|\frac{a-r_-}{b-r_-}\right|^{\frac{\kappa_+}{2\kappa_-}}\left(\frac{a+r_++r_-+\delta}{b+r_++r_-+\delta}\right)^{\frac{(r_++r_-+\delta)^2(r_+-r_-)}{2r^2_+(2r_-+r_++\delta)}},
\end{eqnarray} 
where we have used the fact that $t_a=t_b$ extremizes the generalized entropy. In order to find the spatial location of the island boundaries, we extremize the generalized entropy $S^{\text{lt}}_{\text{gen}}$ with respect to $a$ where the boundaries of the island are located in the vicinity of the event horizon. The solution for $a=r_h+\epsilon+\mathcal{O}(\epsilon^2)$ to the subleading order approximation is derived as
\begin{eqnarray}
a\simeq r_+\left[1+\left(\frac{cG_N}{12\pi r^2_+}\right)^2\frac{r_+}{(b-r_+)}\left(\frac{r_+-r_-}{b-r_-}\right)^{\frac{\kappa_+}{\kappa_-}}\left(\frac{2r_++r_-+\delta}{b+r_++r_-+\delta}\right)^{\frac{(r_++r_-+\delta)^2(r_+-r_-)}{r^2_+(2r_-+r_++\delta)}}\right].
\end{eqnarray}
Here, we see that in analogy to the case of the neutral black hole the subleading term of the location of the island boundaries is suppressed by $(cG_N/r^2_+)^2$ and the higher order terms should be strongly suppressed by the further powers of $cG_N/r^2_+$. Then, by inserting the location of the island boundaries just obtained into $S^{\text{lt}}_{\text{gen}}$, the entanglement entropy for the non-extremal black hole is finally determined as
\begin{eqnarray}
S_{EE}&=&\frac{2\pi r^2_+}{G_N}+\frac{c}{3}\log\left[\frac{m^2_gc_1}{2}\frac{(b-r_+)(b-r_-)(b+r_++r_-+\delta)}{\kappa^2_+r_+b}\right]+\frac{c}{6}\left(\frac{\kappa_+}{\kappa_-}-1\right)\log\left(\frac{b-r_-}{r_+-r_-}\right)\nonumber\\
&&+\frac{c}{6}\left[\frac{(r_++r_-+\delta)^2(r_+-r_-)}{r^2_+(2r_-+r_++\delta)}-1\right]\log\left(\frac{b+r_++r_-+\delta}{2r_++r_-+\delta}\right)+\mathcal{O}\left(\frac{cG_N}{r^2_+}\right).
\end{eqnarray}
This result indicates that the entanglement entropy at the late times would reach a saturation value whose leading order is the twice of the Bekenstein-Hawking entropy of the black hole, i.e. $S_{EE}\simeq2\pi r^2_+/G_N=2S_{\text{BH}}$. Beside, it obtains the logarithm corrections coming form the quantum matter. Other correction terms are suppressed by the powers of $cG_N/r^2_+$.

We expand the leading contribution of the entanglement entropy $S_{EE}$ for the non-extremal black hole in terms of the graviton mass as
\begin{eqnarray}
S_{EE}=\frac{2\pi r^2_{0+}}{G_N}\left[1-\frac{r_{0+}(2c_2+c_1r_{0+})}{r_{0+}-r_{0-}}m^2_g+\mathcal{O}(m^4_g)\right],
\end{eqnarray}
where $r_{0\pm}=M\pm\sqrt{M^2-Q^2}$ are the the radii of the event and Cauchy horizons of the RN black hole corresponding to Einstein gravity or the case of massless graviton. Similar to the neutral black hole, if the event horizon radius of the corresponding RN black hole (in Einstein gravity) is larger than the ratio $-2c_2/c_1$, the nonzero mass of graviton would reduce the entanglement entropy compared to the massless case. 

The Page time for the non-extremal charged black hole is easily derived as $t_{\text{Page}}=3S_{\text{BH}}/\pi cT_H$ where $T_H=\kappa_+/2\pi$ from equating the linear growth of the entanglement entropy at the early times with the asymptotic constant value $2S_{\text{BH}}$. We expand the Page time in terms of $m^2_g$ as
\begin{eqnarray}
t_{\text{Page}}=\frac{12\pi r^4_{0+}}{cG_N(r_{0+}-r_{0-})}\left[1-\frac{r_{0+}}{2}\frac{4c_2(2r_{0+}-3r_{0-})+c_1r_{0+}(5r_{0+}-7r_{0-})}{(r_{0+}-r_{0-})^2}m^2_g+\mathcal{O}(m^4_g)\right],\label{Pagtime-cBH}
\end{eqnarray}
where the first term is the Page time associated with the RN black hole and the second and higher order terms are the contributions arising due to the nonzero mass of graviton. We observe that if the following relation 
\begin{eqnarray}
\frac{4c_2}{c_1}>-\frac{r_{0+}(5r_{0+}-7r_{0-})}{2r_{0+}-3r_{0-}},\label{mgcps-Paget}
\end{eqnarray}
is satisfied, the entanglement entropy in massive gravity would take a shorter time to reach the saturation value, and on the contrary it would take a longer time for the emergence of the island. As discussed in the case of the neutral black hole, this is because massive gravity with the coupling parameters satisfying (\ref{mgcps-Paget}) reduces the bound of the entanglement entropy or/and increases the Hawking temperature causing the black hole radiating more rapidly. However, there is an important difference between the neutral black hole and the non-extremal charged black hole in massive gravity. With respect to the neutral black hole, the Page time in massive gravity is always smaller than that in Einstein gravity for the coupling parameters $c_{1,2}$ which are both positive. But, with respect to the non-extremal charged black hole, this only happens if $r_{0+}$ (or the black hole mass) is sufficiently far from $r_{0-}$ (or the electric charge of the black hole). Whereas, when $r_{0+}$ is sufficiently close to $r_{0-}$, the second term in (\ref{Pagtime-cBH}) is positive which means that the Page time for the non-extremal charged black hole in massive gravity is larger than that in Einstein gravity. 

We compute the scrambling time for the non-extremal charged black hole in massive gravity, which is obtained as
\begin{eqnarray}
t_{\text{scr}}&\simeq&\frac{1}{2\pi T_H}\log S_{\text{BH}}\nonumber\\
&\simeq&\frac{2r^2_{0+}}{r_{0+}-r_{0-}}\log\left(\frac{\pi r^2_{0+}}{G_N}\right)-\frac{m^2_gr^3_{0+}}{(r_{0+}-r_{0-})^3}\left[2(r_{0+}-r_{0-})(2c_2+c_1r_{0+})\right.\nonumber\\
&&\left.+[4c_2(r_{0+}-2r_{0-})+c_1r_{0+}(3r_{0+}-5r_{0-})]\log\left(\frac{\pi r^2_{0+}}{G_N}\right)\right]+\mathcal{O}(m^4_g).
\end{eqnarray}
This expression implies that the scrambling time is very small compared to the Page time and its behavior in the parameters of massive gravity is basically the same as the Page time for the non-extremal charged black hole discussed above.

In the limit of the vanishing electric charge or the Cauchy horizon radius approaching zero, one can easily see that the size of the island, the entanglement entropy, the Page time, and the scrambling time reproduce the corresponding results for the neutral black hole as obtained in the previous section. Also, in the limit of the massless graviton $m_g\rightarrow0$, our calculations for these quantities would reduce the corresponding computations for the RN black hole reported by the authors in
\cite{LiWang2021}.

\subsection{Extremal case}
\begin{figure}[t]
 \centering
\begin{tabular}{cc}
\includegraphics[width=0.4 \textwidth]{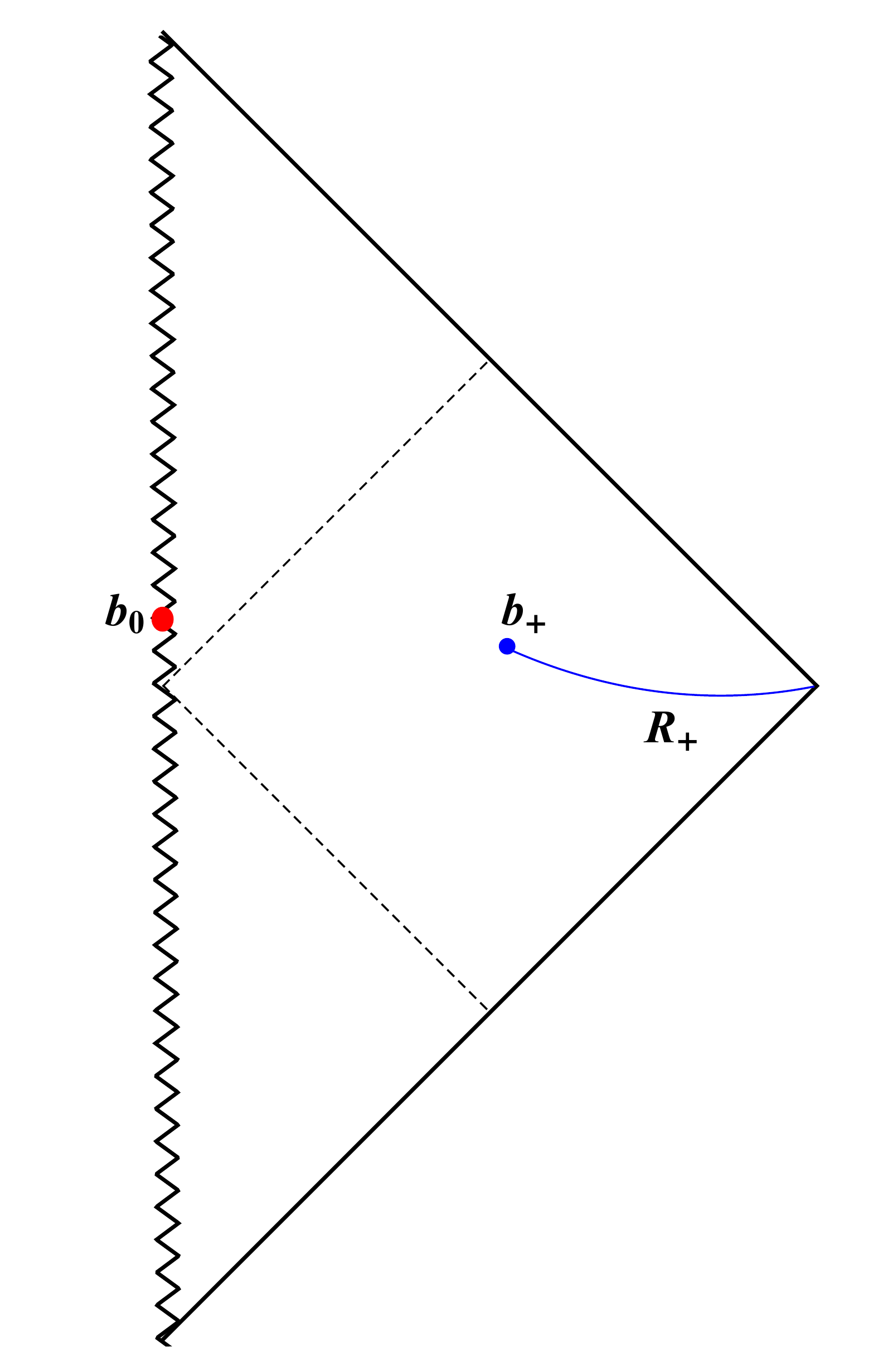}
\hspace*{0.05\textwidth}
\includegraphics[width=0.4 \textwidth]{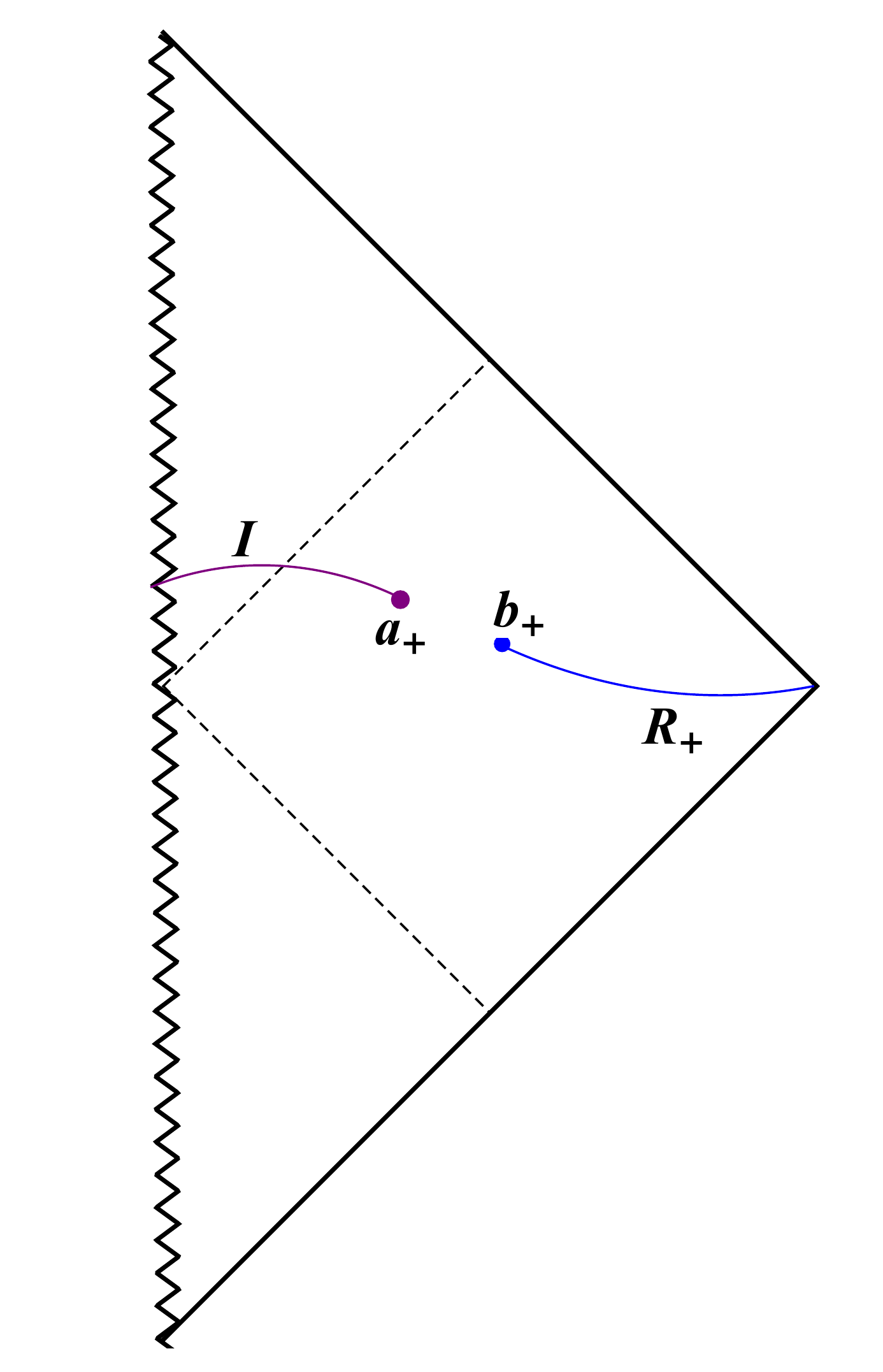}
\end{tabular}
  \caption{The Penrose diagram of the eternal extremal charged black hole in massive gravity without island (left panel) and with one island (right panel). The radiation region is represented by $R_+$ with the boundary $b_+$. $I$ denotes the island extending from the singularity to its boundary $a_+$.}\label{Pd-ex-CBH}
\end{figure}

For the extremal charged black hole, the metric in the Kruskal coordinate is given in Eq. (\ref{Krcood-ECBH}) and the Penrose diagram is shown in Fig. \ref{Pd-ex-CBH}. 
Compared to the non-extremal case, the radiation region is only given by one region $R_+$. In the absence of the island, the entanglement entropy is computed as
\begin{eqnarray}
S(R)&=&\frac{c}{3}\log d(b_+,b_0)\nonumber\\
&=&\frac{c}{12}\log\left[W^2(b)W^2(0)(U(b)-U(0))^2(V(b)-V(0))^2\right],
\end{eqnarray} 
where $b_0=(t_b,0)$ is a reference point which represents the touch of the Cauchy surface at the singularity. Unfortunately, the conformal factor $W^2(0)$ is ill-defined because this function is divergent at $r=0$. As a result, the computation of the entanglement entropy as well as the Page time in the absence of the island would lead to the ill-defined result in the extremal case. However, we hope that the UV corrections at the very short distances such as the nonperturbative renormalization group in the quantization of Einstein gravity \cite{Bonanno2000,Bonanno2006,Koch2016}, string T-duality \cite{Nicolini2019} or noncommutative geometry \cite{Spallucci2009} would make the conformal factor $W^2(r)$ in the Kruskal coordinate behaving smoothly as $r$ approaches zero. And, since it would be able to yield a finite result for the computation of the entanglement entropy and the Page time. This would be further investigated in our future works.

In the presence of an island, the singularity at $r=0$ as discussed above can be avoided because in this situation the entanglement entropy is proportional to the logarithm of the geodesic distance between the boundary of the island and the cutoff surface. More specifically, the generalized entropy for the extremal case with the contribution of one island is computed as
\begin{eqnarray}
S_{\text{gen}}&=&\frac{\pi a^2}{G_N}+\frac{c}{3}\log d(a_+,b_+)\nonumber\\
&=&\frac{\pi a^2}{G_N}+\frac{c}{6}\log\left[\frac{8r^2_e}{m^2_gc_1(3r_e+\delta)^2}\frac{|a-r_e||b-r_e|}{ab}(a+2r_e+\delta)^{\frac{1}{2}}(b+2r_e+\delta)^{\frac{1}{2}}\right]\nonumber\\
&&+\frac{c}{6}\log\left[2\cosh\left(\frac{m^2_gc_1(3r_e+\delta)}{4r_e}(r_*(b)-r_*(a))\right)-2\cosh\left(\frac{m^2_gc_1(3r_e+\delta)}{4r_e}(t_a-t_b)\right)\right],\nonumber\\
\end{eqnarray}
where 
\begin{eqnarray}
\frac{m^2_gc_1(3r_e+\delta)}{4r_e}(r_*(b)-r_*(a))&=&\frac{r_e(b-a)}{2(b-r_e)(a-r_e)}+\frac{5r_e+2\delta}{2(3r_e+\delta)}\log\left|\frac{b-r_e}{a-r_e}\right|\nonumber\\
&&+\frac{(2r_e+\delta)^2}{2r_e(3r_e+\delta)}\log\left(\frac{b+2r_e+\delta}{a+2r_e+\delta}\right).
\end{eqnarray}
It is straightforward to see that the extremal condition $\partial S_{\text{gen}}/\partial t_a=0$ leads to $t_a=t_b$. In addition, we assume that the location of the island boundary is slightly outside the horizon $r_e$ of the black hole, i.e. $a\approx r_e$, as well as the cutoff surface of the radiation region is far away from the horizon $r_e$ of the black hole. With this assumption $b\gg r_e\approx a$, we have the following approximation
\begin{eqnarray}
2\cosh\left(\frac{m^2_gc_1(3r_e+\delta)}{4r_e}(r_*(b)-r_*(a))\right)\simeq e^{\frac{m^2_gc_1(3r_e+\delta)}{4r_e}(r_*(b)-r_*(a))}\gg1.
\end{eqnarray}
By extremizing the generalized entropy within this approximation with respect to $a$, we find the location of the island boundary as
\begin{eqnarray}
a\simeq r_e\left[1+\sqrt{\frac{cG_N}{24\pi r^2_e}}\right],
\end{eqnarray}
where the subleading term is suppressed by square root of $cG_N/r^2_e$, rather than the second power as in the cases of the neutral and non-extremal charged black holes. Then, the entanglement entropy reads
\begin{eqnarray}
S_{EE}=\frac{\pi r^2_e}{G_N}+\sqrt{\frac{6\pi r^2_e}{cG_N}}+\sqrt{\frac{\pi c r^2_e}{6G_N}}+\mathcal{O}(c),
\end{eqnarray}
where the terms relating to the first order of $c$ are the logarithm corrections which are small compared to the subleading terms. We observe that the entanglement entropy for the extremal charged black hole becomes a finite constant at the late times of the evaporation, like the neutral and non-extremal charged black holes. However, there here is an important difference that the asymptotic constant value for the entanglement entropy with respect to the extremal charged black hole is approximately the Bekenstein-Hawking entropy (rather than the twice of the Bekenstein-Hawking entropy), i.e. $S_{EE}\simeq\pi r^2_e/G_N=S_{BH}$. This is due to the essential difference in their causal structure: the Penrose diagram of the neutral and non-extremal charged black holes is the two-sided geometry, whereas the Penrose diagram of the extremal charged black hole is the one-sided geometry. This difference also implies that one cannot derive the entanglement entropy for the extremal charged black hole by taking the continuous extremal limit $r_+\rightarrow r_-$ of the non-extremal charged black hole. 
\section{\label{conclu} Conclusion}

There are the arguments \cite{Hgeng2020,HaoGeng2021} which implied that calculating the entanglement entropy of Hawking radiation and the Page curve for black holes using the island method would be considered in the situation of massive graviton. This is one of the reasons that we consider the entanglement entropy of Hawking radiation and the Page curve using the island method in the framework of dRGT massive gravity which is considered as a candidate for a consistent theory of gravity accompanied with a nonzero mass of graviton. Furthermore, an interesting aspect of black holes in the context of dRGT massive gravity is that the behavior of black hole geometries is asymptotically non-flat. For various black hole geometries corresponding to the asymptotically flat behavior, it was pointed to that the island is emerged at the late time of the black hole evaporation where the boundary of the island is very close the black hole horizon and the island extends almost through the whole black hole interior. In this work, we also show the emergence of the island at the late time of the evaporation the same as the cases of the asymptotically flat black hole geometries. This result and the previous confirmations thus support the emergence of the island at the late time of the evaporation as a universal feature of the semiclassical description of the black hole evaporation: the computations based on the semiclassical description of the black hole evaporation can create replica wormholes or the island where the information is stored; since the black hole evaporation follows the Page curve consistent with the unitarity evolution.

More explicitly, we calculated the entanglement entropy of Hawking radiation emitted by the eternal black holes, the corresponding Page curve, and the scrambling time in the context of massive gravity whose free parameters are the graviton mass and two coupling parameters $c_{1,2}$ in four dimensions. In our calculations, we first employ the island rule which was recently obtained by using the quantum extremal surface technique as well as from the gravitational Euclidean path integral using the replica trick. According to the prescription of the quantum extremal surface, the entanglement entropy is derived as the minimum value of the generalized entropy which is a sum of the Bekenstein-Hawking entropy of the island boundaries and the von Neumann entropy of the matter fields on the union of the radiation region and the island. In order to find the minimum value of the generalized entropy, we need to extremize the generalized entropy over all possible boundary surfaces of the island. Second, we use the approximation of the two-dimensional conformal field theory where the contribution of the matter sector to the entanglement entropy is easily computed as the logarithm of the disjoint intervals, when the distance between the cutoff surfaces of the radiation region is sufficiently large compared to the scale of the size of the cutoff surfaces. 

For the neutral and non-extremal black holes, we indicate that the island does not appear at the early times of the black hole evaporation or in other words the no island configuration is the dominant contribution to the entanglement entropy at the early stage of the evaporation. Hence, the entanglement entropy grows linearly with time. However, as the amount of Hawking radiation becomes sufficiently large at the late times, the island emerges slightly outside the event horizon of the black holes and becomes the preferred configuration by which the growth of the entanglement entropy reaches a finite saturation value which is the twice of the Bekenstein-Hawking entropy of the black hole at the leading order. Furthermore, we calculate the Page time and scrambling time which are approximately given by $\frac{3S_{BH}}{\pi cT_H}$ and $\frac{\log S_{BH}}{2\pi T_H}$, respectively, which is universal for various black hole geometries \cite{Hashimoto2020,LiWang2021,KimNam2021,Karananas2021,YuGe2021,Ahn2021,LuLin2021}. Whereas, for the extremal charged black hole, the entanglement entropy at the early times without the island is ill-defined because the conformal factor of the metric in the Kruskal coordinate is divergent at $r=0$ corresponding to that the Cauchy surface hits the curvature singularity. This implies that we need to consider new physics in the UV region which would make the metric behaving smoothly at the very short distances, which will be further investigated in our future works. At the late times when the island is formed, the entanglement entropy for the extremal charged black hole reaches the saturation value which is the Bekenstein-Hawking entropy. The differences between the extremal charged black hole and the non-extremal charged (and neutral) black hole are due to their Penrose diagram: the Penrose diagram of the neutral and non-extremal charged black holes is the two-sided geometry, whereas it is the one-sided geometry for the extremal charged black hole. We also show that the corresponding results of the Schwarzschild and Reissner-Nordstr\"{o}m black holes are restored in the limit of that the graviton mass goes to zero.

In addition, we study the impact of the parameters of massive gravity on the size of the island, the entanglement entropy, the Page time, and the scrambling time. The direct impact of massive gravity can be ignored compared to its indirect impact which modifies the horizon radius of the black hole. An analytic investigation can be performed in the region of the small graviton mass which is consistent with the presently experimental constraint. If the black hole mass $M$ for the case of the zero electric charge (or a combination of the mass and the electric charge of the black hole for the case of the nonzero electric charge) is larger than $-c_2/c_1$, the bound of the entanglement entropy in massive gravity is lower than that in Einstein gravity, and this happens the contrary for $M<-c_2/c_1$. This suggests that, with the nonzero mass of graviton and the proper coupling parameters, it takes a shorter duration in order for the entanglement entropy reaching the saturation value.

\end{document}